\pdfoutput=1
\documentclass[twocolumn]{aastex63}
\usepackage{amsmath,amssymb}
\usepackage{xspace}
\newcommand{\be}{\begin{equation}}
\newcommand{\ee}{\end{equation}}
\newcommand{\bex}{\begin{equation}\notag}
\newcommand{\eex}{\end{equation}\notag}
\newcommand{\bea}{\begin{eqnarray}}
\newcommand{\eea}{\end{eqnarray}}
\newcommand{\beax}{\begin{eqnarray*}}
\newcommand{\eeax}{\end{eqnarray*}}
\newcommand{\ba}{\begin{array}}
\newcommand{\ea}{\end{array}}

\newcommand{\bs}{\boldsymbol} 
\newcommand{\vecB}{{\bs B}}
\newcommand{\vecE}{{\bs E}}

\newcommand{\vecA}{{\bs A}}

\newcommand{\vecV}{{\bs v}}

\newcommand{\unitv}[1]{\mbox{\boldmath$\hat{#1}$}}

\newcommand{\boldv}{\bs v} 
\newcommand{\bb}{\bs B} 
\newcommand{\bolde}{\bs E} 
\newcommand{\bolda}{\bs A} 
\newcommand{\bolds}{\bs S} 
\newcommand{\bcr}{\bs\times} 
\newcommand{\bdel}{\bs\nabla} 
\newcommand{\bdot}{\bs\cdot} 
\newcommand{\delcr}{\bdel\bcr} 
\newcommand{\deldot}{\bdel\bdot} 

\newcommand{\boldb}{\bs B}  

\newcommand{\boldi}{\bs I}
\newcommand{\boldg}{\bs g}
\newcommand{\boldJ}{\bs J}

\newcommand{\radmhd}{\textsl{RADMHD}\xspace}

\received{June 10, 2020}
\revised{}
\accepted{}
\submitjournal{ApJ}

\shorttitle{The CGEM Model}
\shortauthors{Hoeksema et al.}

\begin{document}


\title{The Coronal Global Evolutionary Model: Using HMI Vector Magnetogram and Doppler Data to Determine Coronal Magnetic Field Evolution}

\correspondingauthor{J. Todd Hoeksema}
\email{todd@sun.stanford.edu}

\author[0000-0001-9130-7312]{J. Todd Hoeksema}
\affiliation{W. W. Hansen Experimental Physics Laboratory, 
Stanford University,
Stanford, CA 94305, USA}

\author[0000-0003-3961-2381]{William P. Abbett}
\affiliation{Space Sciences Laboratory,
University of California,
7 Gauss Way,
Berkeley, CA 94720-7450, USA}

\author[0000-0001-5540-8108]{David J. Bercik}
\affiliation{Space Sciences Laboratory,
University of California,
7 Gauss Way,
Berkeley, CA 94720-7450, USA}

\author[0000-0003-2110-9753]{Mark C. M. Cheung}
\affiliation{Lockheed Martin Solar and Astrophysics Laboratory,
Building 252, 3251 Hanover Street,
Palo Alto, CA 94304, USA}

\author[0000-0002-6338-0691]{Marc L. DeRosa}
\affiliation{Lockheed Martin Solar and Astrophysics Laboratory,
Building 252, 3251 Hanover Street,
Palo Alto, CA 94304, USA}

\author[0000-0002-6912-5704]{George H. Fisher}
\affiliation{Space Sciences Laboratory,
University of California,
7 Gauss Way,
Berkeley, CA 94720-7450, USA}

\author[0000-0001-9046-6688]{Keiji Hayashi}
\affiliation{W. W. Hansen Experimental Physics Laboratory, 
Stanford University,
Stanford, CA 94305, USA}
\affiliation{NorthWest Research Associates,
3380 Mitchell Lane,
Boulder, CO 80301-2245}

\author[0000-0001-8975-7605]{Maria D. Kazachenko}
\affiliation{Astrophysical and Planetary Sciences,
University of Colorado,
2000 Colorado Avenue,
Boulder, CO 80309, USA}
\affiliation{Space Sciences Laboratory,
University of California,
7 Gauss Way,
Berkeley, CA 94720-7450, USA}

\author[0000-0002-0671-689X]{Yang Liu}
\affiliation{W. W. Hansen Experimental Physics Laboratory,
Stanford University,
Stanford, CA 94305, USA}

\author[0000-0003-2045-5320]{Erkka Lumme}
\affiliation{Department of Physics, University of Helsinki, 
Helsinki, Finland}

\author[0000-0001-6886-855X]{Benjamin J. Lynch}
\affiliation{Space Sciences Laboratory,
University of California,
7 Gauss Way,
Berkeley, CA 94720-7450, USA}

\author[0000-0003-4043-616X]{Xudong Sun}
\affiliation{Institute for Astronomy,
University of Hawaii at Manoa,
Pukalani, Hawaii 96768, USA}
\affiliation{W. W. Hansen Experimental Physics Laboratory,
Stanford University,
Stanford, CA 94305, USA}

\author[0000-0003-2244-641X]{Brian T. Welsch}
\affiliation{Natural and Applied Sciences,
University of Wisconsin, Green Bay,
Green Bay, WI 54311, USA}
\affiliation{Space Sciences Laboratory,
University of California,
7 Gauss Way,
Berkeley, CA 94720-7450, USA}

\begin{abstract}
The Coronal Global Evolutionary Model (CGEM) provides data-driven simulations of the magnetic field in the solar corona to better understand the build-up of magnetic energy that leads to eruptive events. 
The CGEM project has developed six capabilities. 
CGEM modules (1) prepare time series of full-disk vector magnetic field observations to (2) derive the changing electric field in the solar photosphere over active-region scales. 
This local electric field is (3) incorporated into a surface flux transport model that reconstructs a global electric field that evolves magnetic flux in a consistent way. 
These electric fields drive a (4) 3D spherical magneto-frictional (SMF) model, either at high-resolution over a restricted range of solid angle or at lower resolution over a global domain, to determine the magnetic field and current density in the low corona. 
An SMF-generated initial field above an active region and the evolving electric field at the photosphere are used to drive (5) detailed magneto-hydrodynamic (MHD) simulations of active regions in the low corona. 
SMF or MHD solutions are then used to compute emissivity proxies that can be compared with coronal observations. 
Finally, a lower-resolution SMF magnetic field is used to initialize (6) a global MHD model that is driven by an SMF electric-field time series to simulate the outer corona and heliosphere, ultimately connecting Sun to Earth.
As a demonstration, this report features results of CGEM applied to observations of the evolution of NOAA Active Region 11158 in February 2011.
\end{abstract}


\section{Introduction to the Coronal Global Evolutionary Model} \label{sec:intro}
The existence of full-disk high-resolution vector magnetic field data taken with an uninterrupted cadence of several minutes from instruments such as the Helioseismic and Magnetic Imager (HMI) \citep{Scherrer2012,Hoeksema2014,Schou2012} on NASA's Solar Dynamics Observatory (SDO) \citep{Pesnell2012} motivates us to ask these questions:  Is it possible to use these data to construct a physics-based model for the evolution of the magnetic field in the Sun's atmosphere?  Can such a data-driven model provide useful insight and predictive capability for understanding how magnetic energy builds up in the solar corona before the occurrence of solar flares and coronal mass ejections?  The overall goal of the Coronal Global Evolutionary Model (CGEM) is to provide such a data-driven modeling capability.
This report describes the scope and capabilities of CGEM, a project funded by Strategic Capability grants from NASA's Living with a Star Program and from the National
Science Foundation \citep[][see also \url{http://cgem.ssl.berkeley.edu}] {Fisher2015}.  To illustrate CGEM's capabilities,
this article focuses on using data and simulations of the evolution of NOAA Active Region 11158 to demonstrate the different components of CGEM, show how they are related to one another, and illustrate at a practical level what is involved in using the various component models of CGEM.

A principle objective of CGEM is to develop a spherical version of the magnetofrictional model \citep{Cheung2012} of the solar corona to study the buildup of magnetic energy. 
The spherical magnetofrictional model (SMF) is driven by time series of magnetic and electric fields determined at the solar photosphere from measurements of the photospheric magnetic and velocity fields made by the HMI instrument on NASA's SDO mission.  
The SMF can be run on either active-region scales or on global scales.
Output from the SMF model can then be used as a starting point for more detailed MHD simulations.  

The CGEM project comprises four main science activities:

\begin{enumerate}
\item  Implement enhanced processing of SDO/HMI vector and full-disk 
line-of-sight magnetogram sequences and HMI Doppler velocity measurements 
and make these available to the Solar Physics and Space Weather communities.
\item  Use these data to compute electric fields at the photosphere, 
on both active-region and global scales, and make the electric field 
solutions publicly available.
\item Use the time sequence of photospheric magnetic field and electric 
field maps to drive a time-dependent, non-potential model based on 
magnetofriction for the magnetic field in the coronal volume, both in 
active regions and globally.  This is done in spherical geometry, in 
either spherical wedge configurations for active regions, or for the 
global Sun. Hereafter, the term ``spherical wedge'' refers to a finite 
sub-volume in spherical coordinates, defined by upper and lower limits of 
radius, latitude, and longitude.
\item For unstable configurations discovered with the SMF model, perform 
follow-up studies using MHD models to provide more realistic dynamics for 
erupting magnetic structures.
\end{enumerate}

To complete these activities, we identify six deliverables for CGEM:

\begin{itemize}
\item D1: Develop a local spherical wedge and a global spherical MF
model.
\item D2: Develop a global flux-transport model that includes areas outside of ``CGEM Patch'' regions to augment the vector magnetogram data within CGEM Patch regions.

\item D3: Develop spherical wedge electric field solutions (from HMI data) inside of CGEM Patch regions, and develop the means to interface these solutions with the flux transport model (D2).

\item D4: Incorporate new enhanced data products into the HMI pipeline to automatically generate electric fields, Poynting and helicity fluxes at the photosphere by integrating the electric field (D3) solutions with the other HMI data products.

\item D5: Develop a simplified MHD model to follow unstable active region configurations found in local SMF simulations.

\item D6: Develop and refine global MHD solutions to connect the global Spherical MF (SMF) model with global heliospheric models.
\end{itemize}

The order of the deliverables given above makes sense for the purpose
of describing the scientific requirements, but does not reflect
the order of workflow needed to actually accomplish the necessary
calculations.  
From this perspective, the first task is to perform whatever enhanced data processing (D4) is needed on the input measurements to compute electric fields at the solar photosphere. 
The next task is to compute the electric field solution at the photosphere (D3) in evolving patches of interest.
Then the local electric field patches are inserted into a global surface flux transport (SFT) model (D2) that allows magnetic flux to emerge consistently with the model evolution computed outside of active regions.
The electric field in patches can be used to run the spherical magnetofrictional model (D1) at active-region scales, or the output of the SFT model can be used as input for the global version of SMF.
Finally, the local SMF model output can be used as a starting point for local-scale MHD simulations (D5), 
or the global version of the SMF model can be used to provide input for global scale MHD simulations (D6) that extend into the heliosphere.  

\begin{figure*}[ht!]
\centerline{ \includegraphics[width=5.5in]{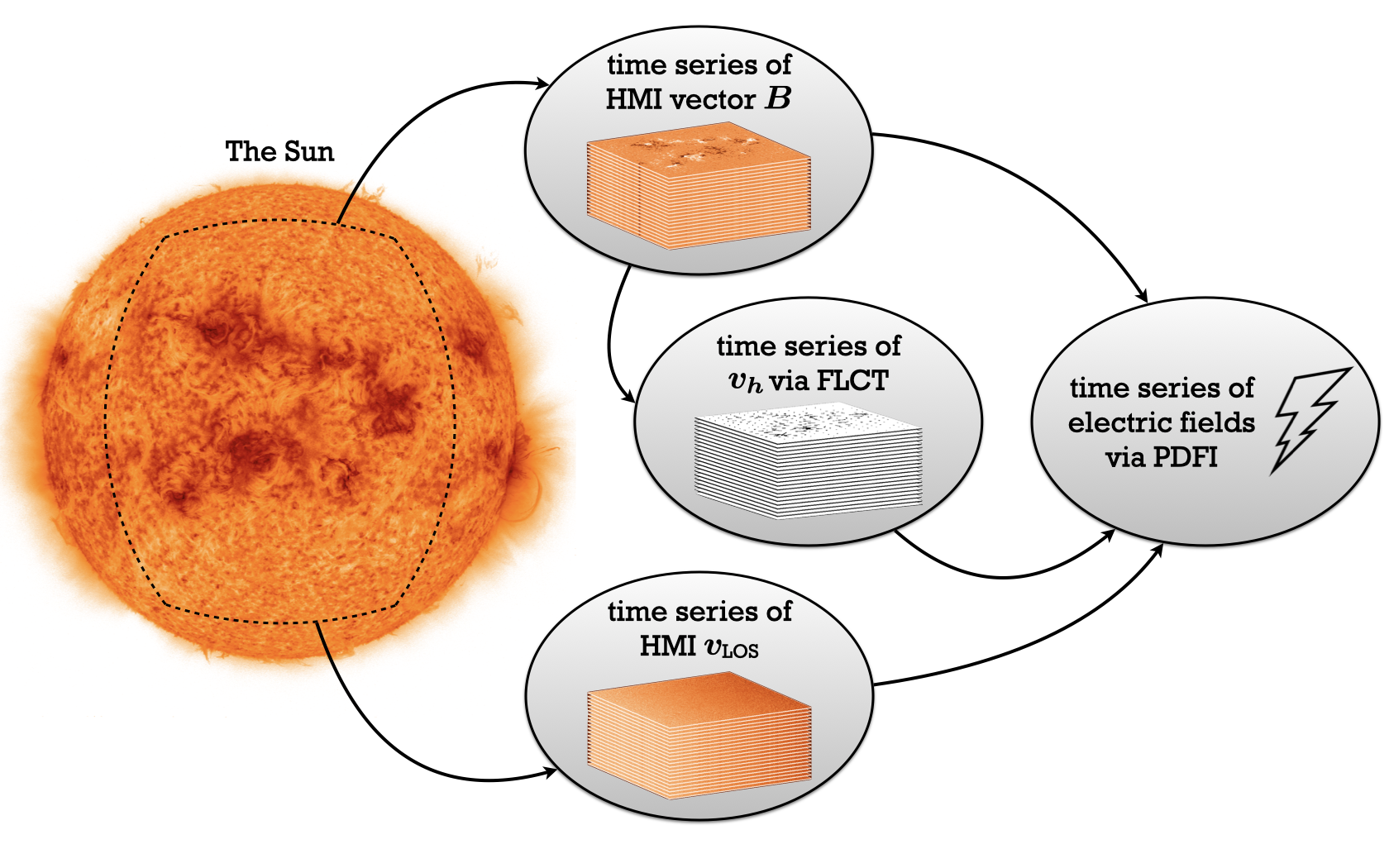} }
\caption{Workflow diagram showing how the D4 deliverable generates enhanced 
data products (corrected HMI vector magnetograms, calibrated Doppler maps, 
and local correlation-tracking velocities) that feed the D3 deliverable 
(electric field inversion software) to produce a time series of photospheric 
electric field maps.  The quantity $\vecB$ is the photospheric vector magnetic field, 
$\vecV_\text{LOS}$ is the observed Doppler velocity, and $\vecV_h$ is the derived horizontal velocity at the photosphere. \label{fig:workflow1}}
\end{figure*}

\begin{figure*}[ht!]
\centerline{ \includegraphics[width=5.5in]{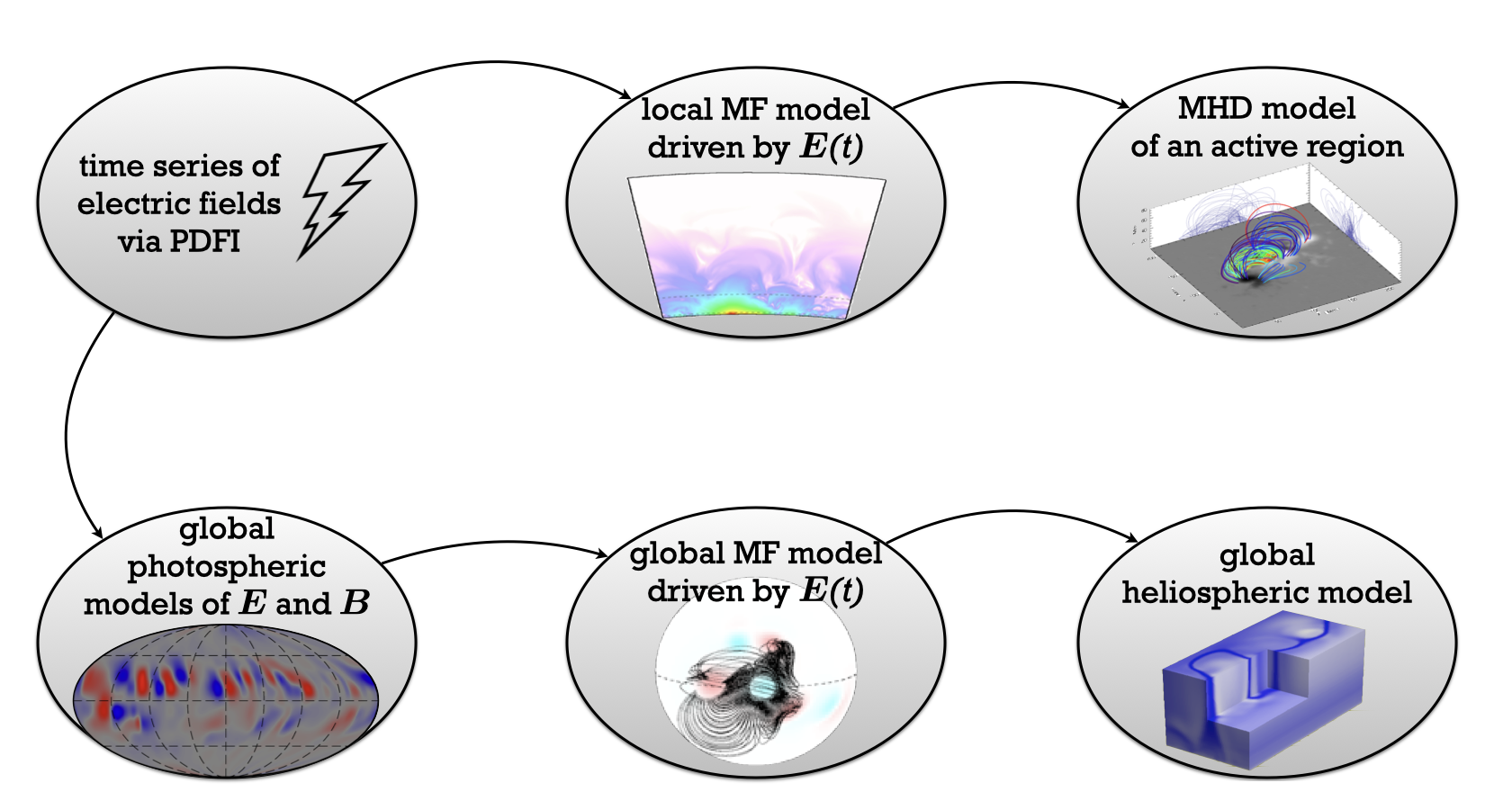} }
\caption{Local electric field maps can be used to drive the local spherical magnetofrictional model (D1)
above active regions, or they can be ingested into the global surface flux transport 
model (D2) that produces quantities used to drive the global SMF model (D1). 
The output from SMF models can then be used as the starting point for MHD simulations
in active regions (D5), or a global coronal and heliospheric MHD model (D6).
\label{fig:workflow2}}
\end{figure*}

The order of topics discussed in this paper follows the workflow requirements.  
To clarify the workflow in an intuitive fashion, we show the workflow order in 
Figures \ref{fig:workflow1} and \ref{fig:workflow2}.


The remainder of the paper is outlined below.

First, in \S\ref{sec:pdfi}, the analysis of the HMI data for NOAA AR~11158 is described (D4).  This is followed by a discussion of the calculation of the electric field solutions at the photosphere (D3).  The data analysis and electric field calculations in general are described in full detail in \citet{Fisher2019}, so the emphasis here is on how the results can be obtained from the Joint Science Operations Center (JSOC) site to be used as input into the other elements of the CGEM model.

The CGEM surface flux transport (SFT, D2) model is described in \S\ref{sec:sftm}.  
A novel aspect of this particular flux transport model is that it is electric-field based, allowing us to more easily interface the model's electric field solutions with those from the smaller-scale active region solutions described in \S\ref{sec:pdfi}.  
Examples showing the global magnetic field configuration during the two months leading up to the CME eruption of 2011 February 12 using the SFT model are shown.

In \S\ref{sec:smfmodel} the spherical magnetofriction model (SMF, D1) is discussed.
Considerable development effort has been completed since the model's initial
description in \citet{Cheung2012}, and the most important of these changes
are described. A simulation of NOAA AR~11158 is performed to both demonstrate
the usage of the model and to show some of the resulting output magnetic configurations. The current status of the model at NASA’s Community Coordinated Modeling Center (CCMC) is summarized.

In \S\ref{sec:driving}, we present the implementation of three-component
data driving for active-region scale MHD simulations (D5), using initial states
from the SMF model and electric field solutions at the photosphere.

\S\ref{sec:emissivity} describes how the current-based
emissivity model developed by \citet{Cheung2012} has been modified to visualize
magnetic field configurations in global spherical geometries.  Both the
advantages and limitations of this software are discussed, along with some
examples of the AR~11158 magnetic configuration shown in a global context.

We discuss the coupling of the global version of the SMF model to our global coronal-heliospheric MHD model (D6) in \S\ref{sec:global}, using the
two months prior to 2011 February 12 to illustrate the coupling of the two
models and the global magnetic field evolution.

Finally, in \S\ref{sec:conclusion}, we summarize the work presented here, then discuss the role of CGEM models in understanding fundamental problems in heliophysics, and suggest directions for future work employing the CGEM models and concepts used in the development of these models.

\section{Advanced HMI Pipeline Processing to Estimate Photospheric Electric Field}
\label{sec:pdfi}

\begin{figure*}[t!]
    \centerline{
    \includegraphics[width=1.0\textwidth]{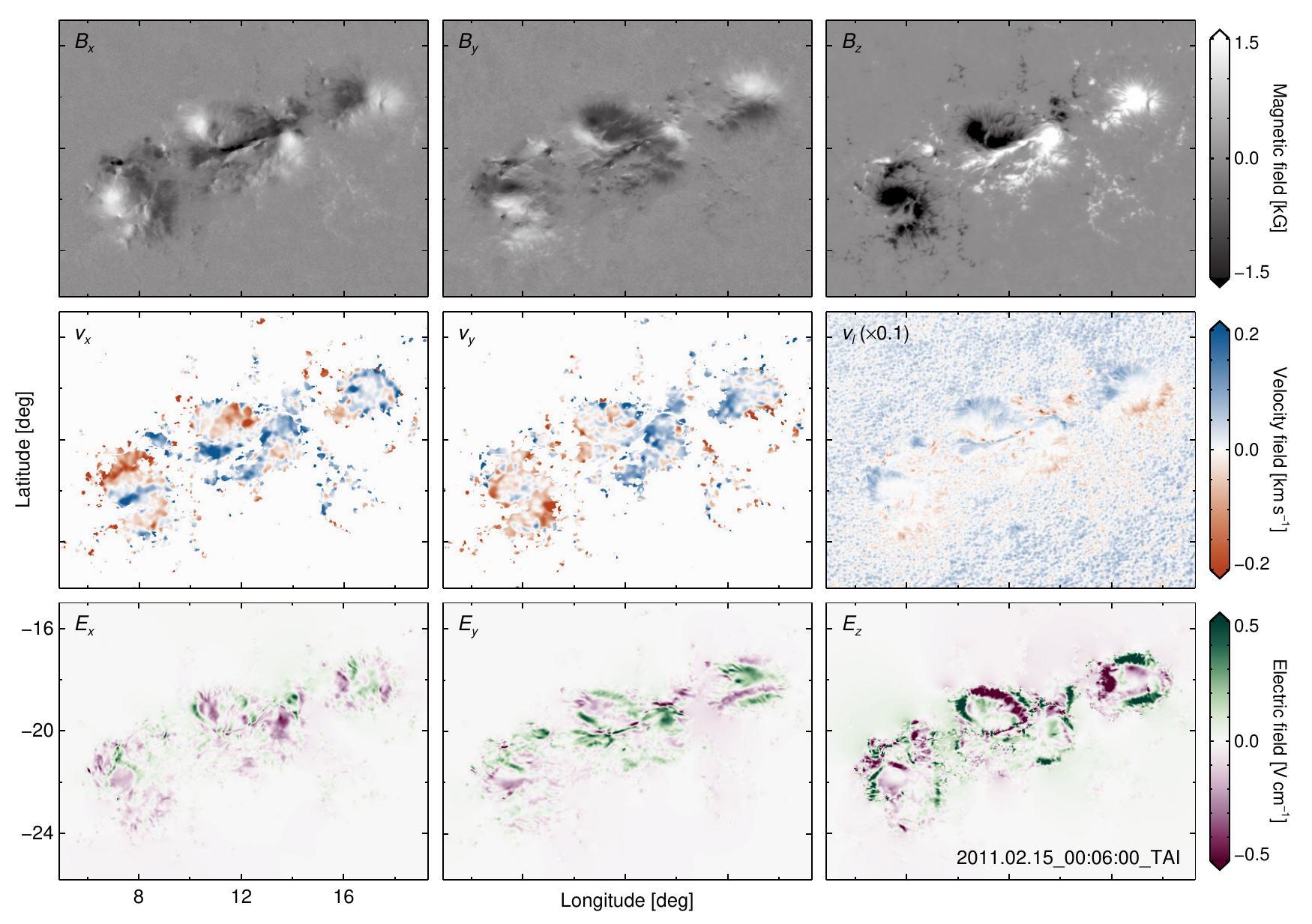}
    }
    \caption{A snapshot of the processed HMI data in AR~11158. The top and middle rows show the vector magnetic and velocity fields. They are the average of the two input frames at 2011.02.15{\_}00:00:00{\_}TAI and 00:12:00{\_}TAI. The lower row shows the inferred electric field at 00:06:00{\_}TAI. Only the central portion of the frame is shown; the weaker field and the padding are excluded. The Doppler velocity plotted here is scaled by 0.1 with blueshift ($+z$) as positive. The original values in the JSOC data set range between $\pm{2}~\mathrm{km~s^{-1}}$ with redshift ($-z$) as positive. For more details see \S\ref{sec:pdfi}. An animation of the entire $6.4$-day data set is available online. }
    \label{fig:pdfi11158}
\end{figure*}

This section presents an overview of the advanced processing of HMI magnetic and velocity data, the electric field inversion method (called PDFI\_SS), 
and ways to access related data products through the SDO Joint Science Operations Center (JSOC). 
For a detailed description of these steps, please consult \S2\,--\,\S4 of \citet{Fisher2019}. Figure~\ref{fig:pdfi11158} shows an example of the output data.

Before deriving electric fields with the PDFI\_SS software, we must first process the full-disk HMI data into a form compatible with PDFI\_SS. Five procedures are necessary to get the data into a suitable form: 
(1) Estimate and remove the Doppler velocity ``convective blue-shift'' bias, due to the overweighting of the hot, upflowing plasma as compared to the cooler, downflowing plasma in the spectral line intensity profile  \citep{Welsch2013}; 
(2) Isolate and track the data centered on an active-region (AR) of interest with a rotation rate defined by the AR center, and map the data into a co-rotating reference frame; 
(3) Correct short-lived azimuth fluctuations in transverse magnetic fields, map the resulting magnetic field, Doppler, and line-of-sight unit vector data into a Plate Carr\'{e}e grid; 
(4) Use successive radial-field magnetograms to estimate apparent horizontal motions using the Fourier Local Correlation Tracking (FLCT) algorithm; and finally,
(5) Add a ribbon of zero-value data around each of the data arrays. We find that this ``zero-padding'' improves the quality of the electric field inversions. As a result of (1)-(5), we derive a final set of vector magnetic and non-orthogonal velocity field components, $(B_x,B_y,B_z)$ and $(v_x,v_y,v_l)$ (see Figure~\ref{fig:pdfi11158}). 
In this section of this article, the subscripts $x$, $y$, and $z$ denote the longitudinal, latitudinal, and radial components of our vectors, respectively. The $l$ subscript denotes the component of a vector projected onto the observer's line of sight, with away from the observer (redshift) positive.  
The source code that performs this calculation is available.\footnote{\url{http://jsoc.stanford.edu/cvs/JSOC/proj/cgem/prep/apps/}} 

We then use PDFI\_SS software to derive the electric field vector, $(E_x,E_y,E_z)$, in the solar photosphere from a time sequence of masked input vector magnetogram and velocity data described above. We construct masks to exclude areas where we expect the noise in the HMI magnetic field to produce unreliable electric fields ($|\vecB|<250$\,G). 
The PDFI\_SS software \citep{Fisher2020pdfi} is based on the PDFI technique for deriving electric fields \citep{Kazachenko2014}. 
The letters in PDFI stand for {\bf P}TD (poloidal-toroidal decomposition), {\bf D}oppler, {\bf F}LCT (Fourier Local Correlation Tracking) and {\bf I}deal, reflecting different contributions into the total electric field as described below. The main idea of the PDFI method is that the electric field can be derived from the observed magnetic field components by un-curling Faraday's law. 
While any such inversion is non-unique due to contributions from gradients of scalar functions, which have zero curl, in PDFI we make use of additional information, in the form of Doppler shifts near polarity inversion lines (PILs), flow-fields from FLCT, and other constraints to compute the electric fields from scalar potentials that can then be added to the solution of Faraday's law to find the total electric field.

Figure~\ref{fig:pdfi11158} shows an example of the processed data described above: the three components of magnetic field, $(B_x,B_y,B_z)$, horizontal and Doppler velocities, $(v_x,v_y,v_l)$, and electric field, $(E_x,E_y,E_z)$, in the central part of AR~11158. 
An online movie
\footnote{\url{http://solarmuri.ssl.berkeley.edu/~kazachenko/public/CGEM_website/pdfi11158.mp4}} 
shows the evolution of these variables during $6.4$ days. The movie shows several interesting phenomena of AR dynamics previously described in \citet{Kazachenko2015} and \citet{ Lumme2019}. The movie also exhibits 
``flickering'' of the electric fields $(E_x,E_y,E_z)$ in the surrounding quiet-sun regions caused by the sensitivity of the PDFI method to noise in the magnetic field and velocity data. We estimate this effect to lead to a $\sim1\%$ error in the overall energy and helicity budgets of the AR (see \S4.2 of \citealt{Lumme2019}).  
Movies made using higher-cadence HMI data ($90$\,s or $120$\,s), which have a lower signal-to-noise ratio than the $12$-minute data, show a much larger amplitude flickering.
 
The PDFI approach has been extensively tested using synthetic data -- magnetograms extracted from MHD simulations where the true photospheric electric field is known \citep{Kazachenko2014}. 
Using anelastic pseudo-spectral ANMHD simulations of an emerging magnetic bipole in a convecting box \citep{Abbett2000,Abbett2004}, we have shown that the PDFI method significantly improves recovery of the simulation's electric field and energy fluxes when compared to the original PTD method of \citet[][see Table 3 of \citealt{Kazachenko2014}]{Fisher2010}. 
The PDFI inversions compare favorably or tend to be more accurate than certain other state-of-the-art velocity inversion methods (e.g., DAVE4VM, see Table 4 in \citealt{Kazachenko2014} and Figures 11 and 12 in \citealt{Schuck2008}). 
Recently we have improved the accuracy of the PDFI numerical method, replacing the original PDFI ``Cartesian Centered'' (CC) grid with a more accurate ``PDFI\_SS'' version discretized on a ``spherical staggered'' (SS) grid \citep{Fisher2019}. 
PDFI\_SS software \citep{Fisher2020pdfi} is written as a general purpose FORTRAN library and can be easily linked to other FORTRAN, C/C++, or Python programs. 
For routine CGEM processing, the SDO/HMI JSOC pipeline software calls one of the high-level Fortran subroutines within PDFI\_SS to compute AR electric fields.

The processed HMI input data sets and the output data sets from PDFI\_SS are publicly available through the SDO JSOC website,\footnote{\url{http://jsoc.stanford.edu}} with series names \verb+cgem.pdfi_input+ and \verb+cgem.pdfi_output+, respectively. 
SDO data analysis manuals contain the details on data query and retrieval methods.\footnote{\url{http://jsoc.stanford.edu/How_toget_data.html}}$^,$\footnote{\url{https://www.lmsal.com/sdouserguide.html}} 
Each JSOC record in the two data series is identified via two keywords, \verb+CGEMNUM+ and \verb+T_REC+.
The keyword \verb+CGEMNUM+ is the NOAA active region number when the CGEM region corresponds to a single named active region, and is equal to 100,000 plus the HMI SHARP number when it does not.  
The keyword \verb+T_REC+ corresponds to the observation time and differs slightly between \verb+cgem.pdfi_input+ and \verb+cgem.pdfi_output+.  
The nominal \verb+T_REC+ is designated at 06, 18, 30, 42, and 54 minutes after the hour for \verb+cgem.pdfi_output+, and at 00, 12, 24, 36 and 48 minutes after the hour for \verb+cgem.pdfi_input+. 
For example, users can find a pair of input records for AR~11158 at the beginning of 2011 February 15 with \verb+cgem.pdfi_input[11158][2011.02.15_00:00-+\\\verb+2011.02.15_00:12]+, which includes the vector magnetic field, FLCT velocity field, Doppler velocity, and local unit normal vectors. 
The corresponding PDFI output can be found with \verb+cgem.pdfi_output[11158]+\\\verb+[2011.02.15_00:06]+, containing vector magnetic fields, electric fields, and the Poynting and helicity fluxes on a staggered grid (for a full list of output variables see \S10.3 of \citealt{Fisher2019}). 

To provide a sense for the computational resources needed to produce electric field solutions, our tests show that using a single processor on a 2016 model Macbook Pro laptop, electric field solutions at a single time are produced in roughly 5\,s given vector magnetogram and Doppler data for the two consecutive time steps of the input data series for AR~11158.

Finally, we wish to point out another data product developed as part of CGEM to understand magnetic activity. 
The JSOC data series \texttt{cgem.lorentz} provides a comprehensive calculation of Lorentz forces in all of the active regions observed by HMI \citep{sun2014}.
For each SHARP region at each time stamp, this data series provides three maps of the photospheric Maxwell stress tensor and their surface integral. 
The latter can be viewed as a proxy for the integrated Lorentz force in the entire volume above the photosphere; it is computed using the divergence theorem and a few simplifying assumptions \citep[see e.g.,][]{Fisher2012a}. 
During eruptive solar events, Lorentz forces computed with HMI are sometimes observed to undergo abrupt changes that coincide in time with the events.
Fast evolution of the photospheric field during major solar eruptions is also clear \citep{sun2017}.
The data series can also be useful for evaluating the ``force-freeness'' of the photospheric field when it is used as the input for coronal field extrapolation \citep{wiegelmann2012}. 

\section{A Global Surface Flux Transport Model Based on the Electric Field}
\label{sec:sftm}

Global surface flux transport (SFT) models describe the evolution of the magnetic field $B_r$ on the photosphere of the Sun. 
SFT models are found to match the observed evolution of the radial component $B_r$ of the photospheric magnetic flux on the Sun reasonably well, and are widely used within the field of solar physics \citep[for additional details, see the review by][]{JiangJ2014}. 
At their core, these models solve the radial component of the magnetic induction equation,

\begin{equation}\label{eq:induction}
    \frac{\partial\bb}{\partial t}=\delcr(\boldv\times\bb)-\delcr(\eta\,\delcr\bb),
\end{equation}

\noindent on a spherical surface using prescriptions for the advection of magnetic flux (via the $\boldv\bcr\bb$ term) and the dispersal of flux (via the diffusion term) across the photosphere. 
These flow patterns are often characterized by the empirically determined advection velocity $\boldv$ and the diffusion coefficient $\eta$. 
New flux is added to a SFT model either via a source function or by assimilating (or directly inserting) observed magnetic field measurements into the model. 

In this section, we describe a different implementation of a global SFT model based on photospheric electric fields. 
Here, the evolution of $B_r$ is governed by the radial component of Faraday's Law,

\begin{equation}\label{eq:faraday}
    \frac{\partial\bb}{\partial t}=-c\delcr\bolde,
\end{equation}

\noindent which describes the evolution of $B_r$ on a spherical surface in terms of the curl of horizontal (i.e., zonal and meridional) electric fields. 
The factor $c$ is the speed of light (cgs units are used throughout this section).

\begin{figure*}
    \centerline{
    \includegraphics[width=1.0\textwidth]{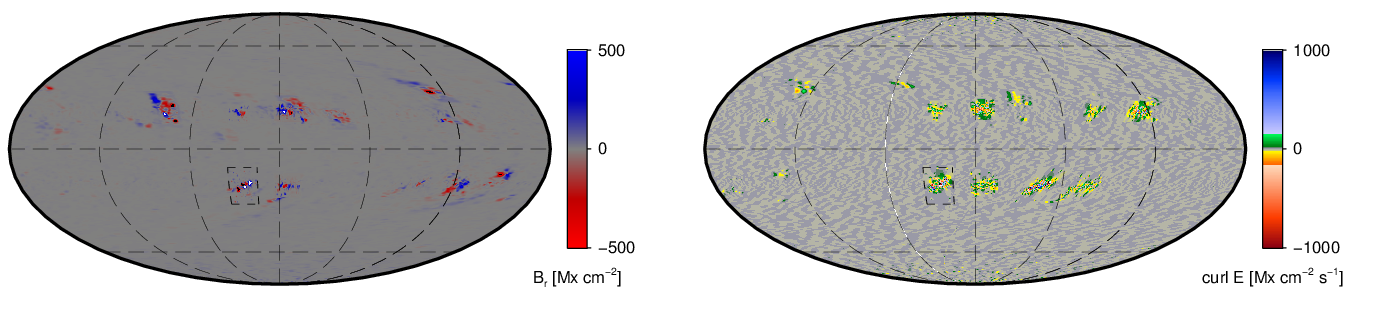}
    }
    \caption{Maps of $B_r$ (left panel) and $\delcr\bolde_h$ (right panel) from the CGEM SFT model for 2011 February 14 at 08:21~UT. PDFI electric fields associated with ARs~11140--11166 from the first two months of 2011 were inserted in the $\bolde_h$ maps, from which the evolution of $B_r$ was determined.  The rectangular boundary of the CGEM Patch associated with AR~11158 is outlined in a dashed line. The maps use a Mollweide projection with grid lines spaced every 60$^\circ$ and is centered on Carrington longitude 60$^\circ$.}
    \label{fig:sft-figure}
\end{figure*}

The PDFI method for processing the HMI Doppler and vector magnetogram data for active regions described in \S\ref{sec:pdfi} results in time series of photospheric electric field data within and around active regions covered by CGEM Patches. 
This series of photospheric electric fields is, by design, consistent with the observed evolution of surface magnetic flux within each CGEM Patch. 
To capture the life cycles of observed active regions on larger scales, including the subsequent dispersal of flux into surrounding quiet-Sun regions, the CGEM SFT model incorporates such time series of electric field data into a global model for the horizontal electric field $\bolde_h$ that describes the evolution of the photospheric radial magnetic field over the entire photosphere.

To obtain a global map of $\bolde_h$, the electric field for locations outside of CGEM Patches is required. 
By adapting the same concepts used in more traditional SFT models, combining equations~(\ref{eq:induction}) and~(\ref{eq:faraday}) indicates that the evolution of $B_r$ is governed by

\begin{equation} \label{eq:sfteh}
    c\bolde_h=-(\boldv_h\bcr B_r\,\bs{\hat{r}})+\eta_h(\delcr B_r\,\bs{\hat{r}}),
\end{equation}

\noindent where $\boldv_h$ represents the empirically determined horizontal flow fields (i.e., differential rotation and meridional flows), $\eta_h$ is a horizontal diffusion coefficient, and $\bs{\hat{r}}$ is the unit vector in the radial direction. 
The CGEM SFT model uses a differential rotation velocity $\boldv_{\text{DR}}$ of the form

\begin{equation}
    \boldv_{\text{DR}}(\theta)=R_{\odot}\,\left[A+B\cos^2(\theta)+C\cos^4(\theta)\right] \bs{\hat{\phi}},
\end{equation}

\noindent where $\theta$ is the heliographic colatitude and $\bs{\hat{\phi}}$ is the unit vector in the longitudinal direction. 
The quantities $(A,B,C)$ are set to $(2.865, -0.405, -0.422) \times $10$^{-6}$~rad~s$^{-1}$ as found by \citet{Komm1993a}, who determined these values by cross-correlating magnetograms on successive days over a time interval of more than 15 years. 
A companion study by \citet{Komm1993b} determined a meridional flow patter+ $\boldv_{\text{MF}}$ having the functional form

\begin{equation}
    \boldv_{\text{MF}}(\theta)=\left[D\sin(2\theta)-E\sin(4\theta)\right] \bs{\hat{\theta}},
\end{equation}

\noindent where $\bs{\hat{\theta}}$ is the unit vector in the colatitudinal direction and $(D,E)=(-12.9,-1.4)$ m~s$^{-1}$. 
Taking the curl of $c\bolde_h$ from equation~(\ref{eq:sfteh}), using a $\boldv_h$ equal to $\boldv_{\text{DR}}+\boldv_{\text{MF}}$ and choosing $\eta_h$ to be 300~km$^2$~s$^{-1}$, gives $\partial B_r/\partial t$ for the CGEM SFT model by equation~(\ref{eq:faraday}). 
A snapshot from the global SFT model is shown in Figure~\ref{fig:sft-figure}.

The use of time series of $\bolde_h$ to drive the evolution of $B_r$ confers several advantages. 

First, knowing $\bolde_h$ enables the evaluation of the Poynting flux $\bolds$ through the photospheric surface to be quantified in the model, since $4\pi\bolds=c\bolde\bcr\bb$. 
In the PDFI scheme described above, the use of both the observed vector magnetic field and the velocity field in constraining $\bolde$ within CGEM Patches also allows $\bolds$ to be constrained within the CGEM Patches, allowing the energetics of the flux-emergence process to be studied more readily. 

Second, it is more straightforward to use time series of $\bolde$ (compared with time series of $\bb$) as a time-evolving lower boundary condition for data-driven models of the overlying coronal dynamics. 
One such data-driven model is described in the next section.

Third, by using $\bolde_h$, the net flux of $B_r$ through the model is always preserved, regardless of the flux (im)balance within each CGEM Patch.\footnote{To wit, by taking the divergence of equation~(\ref{eq:faraday}) one can show that $\partial/\partial t(\deldot\bb)$ vanishes. Consequently, any $\bolde_h$ field, even random noise, will render the net flux of $B_r$ integrated over the full global SFT model unchanged from its initial value of zero.}
In contrast, more traditional SFT models often suffer from flux-imbalance issues, which may be significant when, for example, active regions are not fully contained within the assimilation window. 
One common strategy to deal with such flux-imbalance issues is to subtract off the net imbalance over the whole spherical surface in the model, a treatment that necessarily affects the \textsl{global} distribution of flux in the these models and may cause far-removed neutral lines and other features of interest to shift. 
In the electric-field based scheme presented here, any flux imbalance within a CGEM Patch ends up being balanced by an offsetting amount of flux distributed uniformly within the patch, resulting in the compensatory flux being \textsl{local} to the CGEM Patch (which presumably better matches where such flux is located in reality).

Numerically, the CGEM SFT model is computed on a spherically staggered grid analogous to the grid used in the PDFI electric field determinations of \S\ref{sec:pdfi}. 
The only differences are that the grid spans a full spherical surface instead of a localized CGEM Patch, and that the SFT model has lower resolution by a factor of at least 10 in order to be computationally feasible. 
Using a staggered grid is ideally suited for taking accurate curls of $\bolde_h$, and additionally enables accurate and fast downsampling to occur as long as the downsampling factor is an integral divisor of the original grid dimensions \citep[cf., Figure~8 of][]{Fisher2019}. 
The staggered-grid scheme used here for the purpose of calculating curls is an example of the constrained-transport method, which we believe was first used in an astrophysical setting by \citet{evans1988}. 
Here, we also use the upwind slope-limiting scheme described in equation~(48) of \citet{Stone1992a}, which follows the method developed in \citet{vanleer1977b}.

One issue that arises when inserting the localized PDFI electric fields into a global $\bolde_h$ map is that of a mismatch at the interface between the two electric field domains. 
While the curls of these electric fields respectively yield the desired evolution of $B_r$ both within and outside of each CGEM Patch, taking the curl across the interface can yield spurious values of $\partial B_r/\partial t$. 
This mismatch results because the two types of $\bolde_h$ maps may differ by the gradient of an unknown scalar function and still yield the proper $\partial B_r/\partial t$ within each respective domain; however, there is no guarantee that these will match across the perimeter of the CGEM Patch. 
To address this issue, before inserting a PDFI electric field map into the global map of $\bolde_h$, we add to the PDFI electric field map a curl-free $\bolde_h$, calculated such that the values of $\bolde_h$ of these curl-free fields around the perimeter of the CGEM Patch match the external values determined from the global $\bolde_h$ map corresponding to the large-scale flows. 
This treatment eliminates the mismatch of electric fields around the perimeters of the CGEM Patches without affecting the evolution of $B_r$. 
For additional details, we refer the reader to \S5.1 of \citet{Fisher2019} in which this process is described more fully.

Another issue that materializes when inserting PDFI electric fields into a global $\bolde_h$ map is that the PDFI electric fields only capture the evolution of active regions at times when they are observed.
As a result, any evolution that occurs when the region of interest is not on the Earth-facing side of the Sun, or when data are missing (for example, in daily short intervals during the SDO spacecraft's semiannual eclipse seasons), is not represented. 
During such intervals the ``nudging" scheme described in \S5.2 of \citet{Fisher2019} provides a way to infer a valid $\bolde_h$ that effects a smooth transition between the $B_r$ at one point in time to the $B_r$ at a later time. 
The nudging scheme is used both to bridge data dropouts and to (roughly) approximate the emergence of flux that occurs prior to the active-region flux within a CGEM Patch appearing on the east limb.

Global maps of $\bolde_h$ computed for two separate weeks-long intervals are available in JSOC data series \texttt{cgem.sft\_global\_nlong0300\_noncontiguous}, \texttt{cgem.sft\_global\_nlong0600\_noncontiguous}, and \texttt{cgem.sft\_global\_nlong1200\_noncontiguous}. 
These data series contain global $\bolde_h$ maps at different spatial resolutions, in which the number of grid points spanning the full 360$^\circ$ of longitude is 300, 600, and 1,200 pixels, respectively. 
The time intervals for which global $\bolde_h$ data are available include January and February 2011 (containing regions such as AR~11158) as well as the end of March 2014 (containing regions such as AR~12017). Assuming data from the \texttt{cgem.pdfi\_output} data series are immediately accessible, advancing the global SFT model by one day in time takes approximately 1, 10, or 100 minutes of wall-clock time on a desktop workstation, depending on spatial resolution.

\section{A Data-Driven Spherical Magneto-Frictional Model for Coronal Energy and Helicity}
\label{sec:smfmodel}

The Spherical Magnetofrictional Model (SMF) is an adaptation of the Cartesian MF code~\citep{Cheung2012,Cheung2015b}. It advances Faraday's induction equation

\begin{equation}
\frac{\partial \bolda}{\partial t} = -c\bolde,\label{eqn:faraday}
\end{equation}

\noindent where $\bolda$ is the vector potential and $\bolde = -c^{-1}\vecV \times \bb$ is the electric field.
\citet{Toriumi2020} performed tests of a number of data-driven models (MF and MHD) against a ground-truth MHD model of flux emergence and found the $\bolde$-field driven MF model to quantitatively reproduce the magnetic field energy and relative helicity in the corona. 

The CGEM SMF code solves the induction equation on a spherical coordinate system consisting of computational cells defined on an $(r,\phi,\theta)$ grid, where $r$ is the radial distance from the solar center and $\phi$ and $\theta$ are the longitudinal and latitudinal coordinates, respectively. 
Like the original Cartesian version, the code uses a staggered grid such that 
\begin{itemize}
\item the vector potential $\bolda$, the electric field $\bolde$, and the current density $\boldJ = c(4\pi)^{-1} \delcr \bb$ are defined on cell edges,
\item the magnetic field $\bb = \delcr{\bolda}$ is defined on cell faces, and
\item the MF velocity $\vecV = \nu c^{-1}\boldJ \times \bb$ is defined at cell corners. 
\end{itemize}
\noindent The magnetofrictional coefficient is $\nu = \nu_0 B^2$, where $\nu_0 = 8\times 10^{-7}$ km$^2$\,s$^{-1}$.
\citet{Cheung2012} chose a height-dependent profile for $\nu_0$ such that the coefficient tapered to zero at the photosphere, motivated by comments regarding the nature of MF evolution by \citet{Low2010}. 
The latter point out that under a line-tied scenario (i.e., $E_\phi = E_\theta = 0$ at $r=R_\odot$) in which the photospheric radial flux distribution does not change, MF evolution will create tangential discontinuities, leading to magnetic reconnection near the photospheric boundary which violates line-tying. 
In a data-driven model, the aim is to continuously drive the bottom boundary based on the observed evolution (i.e., not line-tying) regardless of how the model coronal field behaves. 
Since the MF velocity is never used at the bottom boundary, there is no need to taper $\nu_0$ to zero.
The SMF model is driven at the bottom boundary by setting $E_\phi|_{r=R_\odot}$ and $E_\theta|_{r=R_\odot}$, which is supplied either by PDFI inversions (\S\ref{sec:pdfi}), or by the CGEM SFT model (\S\ref{sec:sftm}). 

\begin{figure*}
\centering
\includegraphics[width=0.9\textwidth]{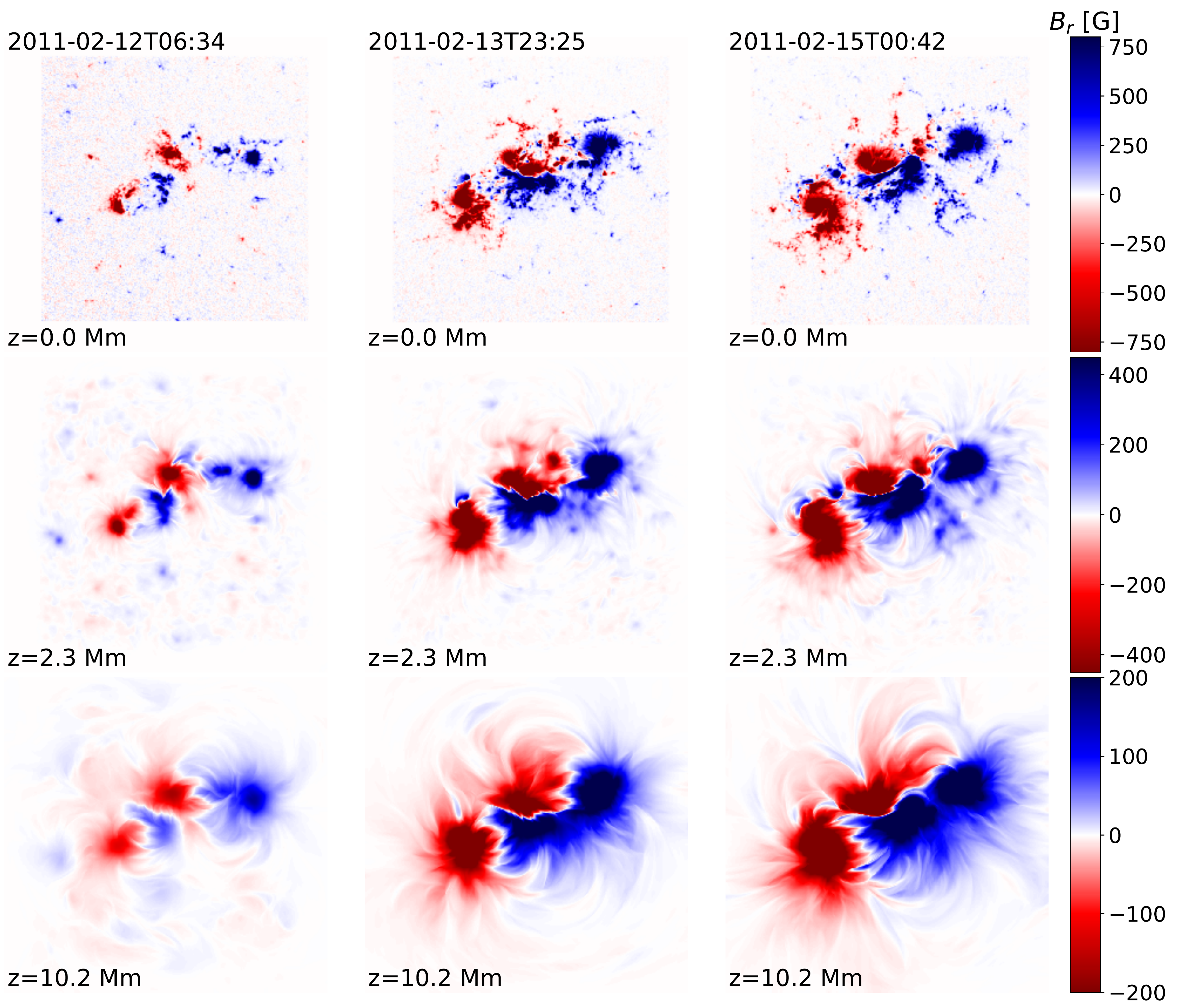}
\caption{Synthetic magnetograms ($B_r$) from the CGEM Spherical Magnetofrictional (SMF) model of AR~11158; columns are different times. Blue and red denote positive and negative values of $B_r$; color bars in each row show scale at each height. Note that the color maps are saturated, $i.e.$ the color map range shown is smaller than the actual range of magnetic field values.}\label{fig:AR11158_SMF}
\end{figure*}

Figure~\ref{fig:AR11158_SMF} shows the radial component of the magnetic field ($B_r$) at different heights $z=r-R_\odot$ in the SMF coronal field model of AR~11158. 
The evolution of the coronal field was driven by CGEM PDFI electric field inversions spanning the 6.4-day time interval \texttt{2011-02-10T14:18:00} to \texttt{2011-02-16T23:42:00}.
To prepare an initial condition for $\bolda$ in the computational volume, $\bolda$ was set to zero for $r\ge R_\odot$. 
In this state, there is a mismatch between the HMI-measured $B_r(r=R_\odot)$ and the model field. 
We use the ``nudging'' technique described in \S5.2 of \citet{Fisher2019} to solve for a correction to the transverse components of the vector potential $(A_\phi,~A_\theta $ at $ r=R_\odot$), such that $(\delcr \bolda)\bdot {\bs{\hat{r}}} = B_r$ at $r=R_\odot$.  
The nudging is is essentially the same as subroutine \texttt{enudge\_ll} described in that article, but with the source code incorporated into the SMF model directly, rather than by linking to the PDFI\_SS library. 
$\bolda$ in the computational volume is then iteratively updated by advancing Eq.~(\ref{eqn:faraday}) with the MF method. 
This relaxation method results in a coronal field that is matched to the initial photospheric flux distribution at \texttt{2011-02-10T14:18:00}.

For the evolving SMF run, the transverse components of the electric field are set by electric field solutions from the PDFI inversion (available via the JSOC data series {\texttt{cgem.pdfi\_output}}). 
By construction the PDFI electric fields are consistent with the observed evolution of $\boldb$ as measured by HMI. 
Imposing this set of electric fields at the bottom boundary of the SMF model drives the evolution of the coronal field in response to the observed photospheric evolution. 
The top and side boundary conditions are implemented such that the field crossing the boundaries is normal to the surface, and the MF-computed velocity field is extrapolated out (zero gradient) into the ghost cells. 
Using a Cartesian version of the MF model to simulate AR~11158, \citet{Chintzoglou2019} showed that the E-field driven coronal field generated a twisted flux rope hours prior to the time of the observed X2-flare at \texttt{2011-02-15T01:43:00}. 
The production of a twisted flux rope also occurs in the spherical MF model.

The complete 6.4-day computational run at the original spatial sampling of the PDFI electric field (as described in this section) required approximately 200,000 CPU hours. 
The SMF module has been delivered to NASA's CCMC and is being implemented for use by researchers.

\section{A Data-Driven Active Region-Scale Radiative MHD Model}
\label{sec:driving}

\subsection{Physics of the Model}

\citet{Abbett2007} developed \radmhd, one of the first numerical models capable of evolving magnetic fields over the vast range of physical conditions and the disparate spatial and temporal scales characteristic of the convection zone-to-corona system.  
Since its initial description in that article, \radmhd has undergone significant updates to improve its ability to model active region evolution in a global environment \citep{Abbett2012,Abbett2014}. 
\radmhd now has the option to evolve the following MHD system of conservation equations on a Cartesian or a spherical-polar block non-uniform mesh, either globally or over a subset of solid angle:

\begin{equation}
  \frac{\partial\rho}{\partial t}+\deldot\left(\rho\boldv\right)=0
  \label{mass}
\end{equation}
\begin{equation}
  \frac{\partial\rho\boldv}{\partial t}+\deldot\left[\rho\boldv\boldv
     +\left(p+\frac{B^2}{8\pi}\right)\boldi-\frac{\boldb\boldb}{4\pi}
     -{\bs\Pi}\right]=\rho\boldg
  \label{momentum}
\end{equation}
\begin{equation}
  \frac{\partial\boldb}{\partial t}+c\,\delcr\bolde=0
  \label{faraday}
\end{equation}
\begin{equation}
  \frac{\partial e}{\partial t}+\deldot\left(e\boldv\right)
    =-p\,\deldot\boldv+\frac{4\pi\eta}{c^2}J^2+\Phi+Q
  \label{energy}
\end{equation}  

\noindent Here, $\rho$, $\boldv$, $\boldb$, $\boldg$, $e$, $p$, $\eta$, and $\bolde= -c^{-1}\boldv\bcr \boldb+4\pi\eta c^{-2}\boldJ$ 
have the standard definitions of gas density, vector velocity, vector magnetic field, local gravitational acceleration, internal energy density, gas pressure, magnetic diffusivity, and vector electric field
(here, the MHD expression for the electric field includes non-ideal processes).  
The current density is expressed in terms of the magnetic field as
$\boldJ=c(4\pi)^{-1}\,\delcr\boldb$, and the system is closed using a tabular
equation of state \citep{Rogers2000} that takes into account the effects of a
partially-ionized gas when relating the internal energy density to the gas
pressure and temperature. 
Within the divergence term of the momentum conservation equation (Eq.~\ref{momentum}), $\boldi$ denotes the identity tensor, and ${\bs\Pi}$ represents the viscous stress tensor for a Newtonian fluid.  
In Eq.~\ref{energy}, $\Phi$ represents the dissipation rate of internal energy through viscous diffusion.

The energy source terms $Q$ are an important component of solar models --- the divergence of the radiative flux near the visible surface drives convective turbulence in surface convection zone-to-corona models, and the interaction of optically thin cooling in the model's corona with the effects of field-aligned electron thermal conduction and Joule heating sets the energy balance and subsequent emission in coronal loops. 
Specifically, we use the  technique of \citet{Abbett2012} to approximate the solution to the gray (frequency-independent) radiative transfer equation in local thermodynamic equilibrium (LTE) assuming a localized, plane-parallel geometry.  
In optically thin regions, the radiative cooling function is expressed as $Q_r=-n_en_h\Lambda(T)$.  
The radiative cooling curve $\Lambda(T)$ is specified using the CHIANTI atomic database \citep{Young2003}. 
The electron and hydrogen number densities $n_e$ and $n_h$ are expressed in terms of the gas density and mean molecular weight as described in \citet{Abbett2007}.  
To include the effects of electron thermal conduction, we employ a field-aligned Spitzer-type conductivity of the form $Q_c=\unitv{b}\,\bdot\bdel(\kappa_{||}(T)\,\unitv{b} \bdot\bdel T)$, where $\unitv{b}$ refers to a local magnetic field-aligned unit vector.  
To mitigate restrictive temperature scale heights characteristic of Spitzer-like conductivity in a model transition region \citep[see e.g.,][where these scale heights can be of order 1~km]{Abbett1999}, 
we implement the adjustments to the temperature-dependent coefficient of thermal conductivity introduced by \cite{Mok2005,Lionello2001,Linker2001}, and used in \citet{Abbett2007}. 
This technique spreads the transition region over a somewhat larger length scale along a magnetic field line and maintains (in an average sense) the equilibrium between thermal conduction and radiative losses in coronal loops.

While other non-local and non-thermal processes (such as Pedersen currents due to cross-field diffusion) can affect the evolution of the model's chromosphere and interface region \citep{Goodman2012, Leake2012, MartinezSykora2013}, 
we neglect these processes here for computational efficacy. 
Our goal is to (1) treat the energetics of the system with sufficient realism over the spatial scales necessary to investigate the interaction of small-scale dynamics with larger-scale magnetic structures typical of active regions, 
and (2) couple dynamics at different scales within the highly-stratified thermodynamic transition between the high-$\beta$ convective interior and low-$\beta$ atmosphere (here, $\beta$ refers to the ratio of gas to magnetic pressure). 

\subsection{Numerical Techniques of the Model}
\label{subsec:techniques}

The \radmhd code\footnote{\url{http://solartheory.ssl.berkeley.edu/cgi-bin/radmhd}} solves the MHD system of equations semi-implicitly using a high-order non-dimensionally split finite-volume formalism that captures and evolves spatial discontinuities.  
The explicit sub-step of the numerical method for the Cartesian case extends the semi-discrete scheme of \citet{Kurganov2000} to three spatial dimensions. 
For the spherical case it extends the 2D curvilinear shock capture scheme of \citet{Illenseer2009} to 3D, 
while simultaneously accounting for area and volume changes in the calculation of numerical fluxes.  
Fluxes are determined using a high-order, 3D conservative, piece-wise continuous interpolating polynomial. 
This allows flows and shocks to be propagated more accurately in off-axis directions.  
A high-order Gaussian integration is used when integrating fluxes over a control volume to update cell averages.

To allow for the incorporation of PDFI electric fields directly into the \radmhd model photosphere in such a way as to be numerically stable and physically self-consistent, 
we implemented the constrained transport method of \citet{Kissmann2012} (extended to 3D curvilinear geometries when using spherical coordinates).  
This scheme is formulated to ensure that electric fields at face edges are consistent between cell volumes that share an edge,
thereby maintaining the solenoidal constraint on the magnetic field to machine roundoff, 
and allowing us to assimilate PDFI electric fields directly into our numerical scheme without introducing additional interpolation error.  
Energy source terms and the effects of viscous stress and magnetic diffusion are treated in the implicit sub-step using an efficient Jacobian-free Newton-Krylov solver \citep[see][]{Knoll2004,Abbett2007}.

\begin{figure*}[t]
    \centerline{ \includegraphics[width=1.0\textwidth]{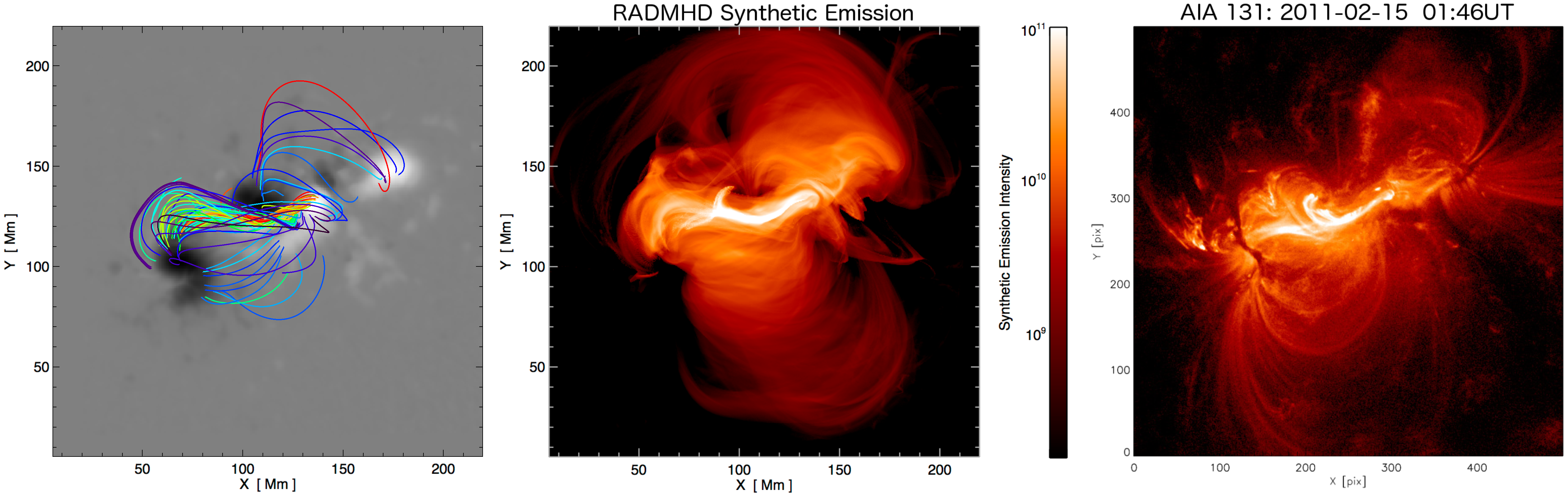} }
    \caption{\footnotesize \radmhd simulation results in the early stages of a PDFI data-driven simulation of NOAA AR~11158 after $\sim$10 minutes of solar time. 
    The simulation's magnetic field was initialized using the Cartesian magnetofrictional model of \citet{Cheung2012}.
    The left panel shows representative magnetic field lines, the center panel shows the synthetic, current-based EUV emission proxy from the \radmhd simulation, and the right panel shows SDO AIA 131\AA\ emission from AR~11158 on 15 February 2011.
\label{fig:j2AIA}}
\end{figure*}

The computational resource requirement for a given \radmhd model entirely depends on one's strategy for the block structure, the resolution required, the physics involved, and the details of the distributed or shared memory computing platform. As a particular example, the pilot simulation shown in Figure~\ref{fig:j2AIA} is of modest scale and was performed on 196 cores of a local distributed memory Intel-based cluster, and took roughly 30 hours of wall-clock time. Each core evolved a block with $64^3$ mesh elements --- the domain decomposition strategy in this case was to minimize inter-processor communication while maximizing the spatial extent of the computational domain. In general, the amount of wall-clock time required for a simulation to run for a given interval of solar time is governed by Courants-Friedrichs-Levy (CFL) stability constraints in the explicit sub-step of the calculation. The size of the time steps in our data-driven simulations are typically determined by the fast magnetosonic wave speeds of the model's low-density corona.

\subsection{Progress on Data-driven MHD Models}
\label{subsec:MHDProgress}

The objectives of our active region-scale data-driven MHD modeling are to 
(1) assimilate PDFI electric fields directly into the photospheric layers of a radiative MHD simulation whose domain encompasses the  highly-stratified transition between the photosphere and low corona, 
and (2) use PDFI electric fields to drive not just the radial component of the model's photospheric magnetic field, but all three components of the field in this active layer. 

We position our model photosphere midway between radial faces of the first active layer of voxels at the base of the computational domain (to be clear, the data-driving here is not imposed via a boundary condition, rather it is done by assimilating data into active zones within the computational domain). 
Therefore, to drive the system, the electric field must be specified along each of the cell edges.  
Yet PDFI data is inherently 2D, and is necessarily limited to the photospheric midplane.
Fortuitously, the PDFI formalism allows us to calculate radial gradients of the angular components of the inductive electric field via Equation 15 of \citet{Fisher2020}. 

With this additional information, we are able to drive all three components of the MHD model's photospheric magnetic field in a physically self-consistent fashion.
%
%
Figure~\ref{fig:j2AIA} shows representative field lines and synthetic emission from a pilot \radmhd data-driven simulation initialized by a Cartesian magnetofrictional state. Yet as these simulations progress, unphysical dynamic behavior can arise in the model's low atmosphere (we describe this behavior in detail later in this section).

\begin{figure*}[!th]
    \centerline{ \includegraphics[width=5.25in]{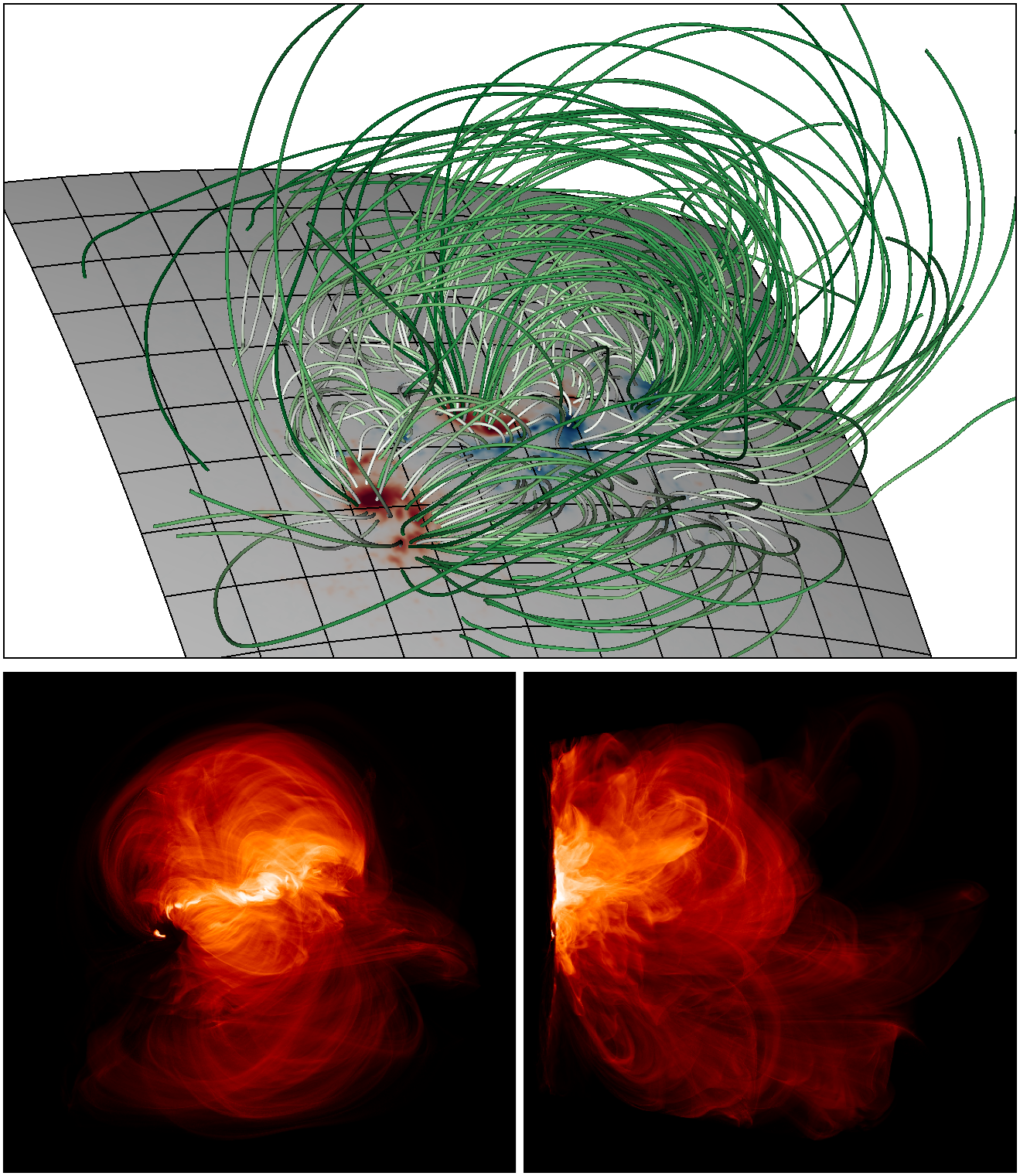} }
    \caption{\footnotesize Magnetic field lines from an SMF model of NOAA AR~11158 imported into the \radmhd domain-decomposed grid for use as an initial state of an MHD simulation of solar activity (\emph{top panel}). 
    Block boundaries are shown at the model's photosphere (black lines). 
    The computational domain spans $20.85^{\circ} \times 20.85^{\circ}$ in the non-radial directions (which corresponds to approximately 253 Mm $\times$ 253 Mm at the photosphere).  
    Each individual block spans $1.74^{\circ} \times 1.74^{\circ}$ (or $\sim$21 Mm in each angular direction at the surface). 
    The current-based emissivity proxy of \S\ref{sec:emissivity} is used to generate synthetic emission at two viewing angles: a view from overhead (\emph{bottom-left panel}) and a view at the limb (\emph{bottom-right panel}).
\label{fig:j2ms}}
\end{figure*}

In the new, spherical \radmhd treatment, our initial magnetic configuration is provided by a pre-eruptive magnetic state, this time generated by the SMF model of $\S$\ref{sec:smfmodel}.
The top panel of Figure~\ref{fig:j2ms} shows $B_r$ on the lower boundary and magnetic field lines from the spherical \radmhd simulation data.  
Knowing the initial magnetic field allows us to calculate an initial driving electric field at the photosphere by taking the difference of the SMF magnetic field and the field specified by the first HMI magnetogram.
We recalculate the initial PDFI electric field instead of using a published CGEM electric field because unless one uses an SMF snapshot exactly corresponding to a magnetogram time, the CGEM field would drive the magnetic field to an increasingly divergent state the farther the SMF snapshot is from a magnetogram time.  
The nudging procedure of \S\ref{sec:smfmodel} is also applied to keep the magnetic field from drifting away from the desired results due to any driving discrepancies that may arise from differences in the numerical methods used in PDFI and those of the \radmhd code. 

To date, much of our effort driving MHD simulations with observational data has been focused on obtaining meaningful comparisons between the CGEM SMF models and MHD models in the zero-$\beta$ limit (an MHD approximation where the effects of gas pressure and gravitational stratification are essentially ignored). 
The advantage of this simplification is that the coronal magnetic field can be efficiently evolved over long periods of time, and the resulting evolution admits to a more direct comparison with existing SMF models.  
There is a significant disadvantage however; namely, the approximation breaks down in the model's photosphere and low atmosphere where strong magnetic fields become concentrated and constrained by gas pressure. 

The SMF model avoids this problem by damping the contributions of its approximate Lorentz force in regions at and above the model's lower photospheric boundary.
A similar approach can be utilized in MHD models (in the zero-$\beta$ limit or otherwise); 
however, the specification of the initial state becomes a significant challenge.
The initial magnetic and thermodynamic configuration (i.e., densities, pressures, and temperatures imposed on the system) is critical to the initial force balance of the system, particularly in the layers at and directly above the model photosphere.  
Unless this initial state is constructed in a physical way, where the pressure gradients act to balance the forces from magnetic pressures and stresses acting to push apart concentrations of field, 
non-physical flows and dynamics will eventually dominate any driving forces imposed at the photosphere. 
This is a particular problem with the zero-$\beta$ approach since the only possible restorative forces in these regions are due to Reynolds and viscous stresses.  

On the Sun, the observed evolution of the magnetic field in the photosphere and low atmosphere results from a complex interaction of fields and flows in a gravitationally stratified, turbulent environment. 
But there is insufficient observational information available to adequately specify the dynamic and thermodynamic initial state of the system.  
In standard {\em ab initio} models of magnetic flux emergence or magnetoconvection one is not faced with this difficulty.  
Typically, field-free thermodynamic states are developed in a physical way through a dynamic and energetic relaxation process, 
and stratification in density, pressure, and temperature naturally results from the presence of gravity and the application of physical boundary conditions. 
Once this relaxation procedure is complete, only then is magnetic field introduced into the system.  

In the case of data driving, we are presented with the opposite scenario. 
We are given an initial magnetic field and an electric field to be applied to the model's photospheric boundary, with little to no information regarding the hydrodynamic state of the plasma in the low atmosphere where such information is necessary to drive the dynamics of the system in a physical way.  

So how best to proceed?  
We find that there are two ways to address this initialization problem: 
The first is to simply scale the fields and place the lower boundary of the simulation in the upper transition region or corona,
thereby placing the driving layer in a field-filled, magnetically-dominated regime.  
The second is to keep the driving layer in the photosphere, generate an initial thermodynamic stratification, 
and artificially limit non-physical runaway flows when necessary to mitigate dynamics unrelated to the driving forces of interest.  

We choose not to pursue the first approach, since we find that photospheric electric fields bear little resemblance to coronal electric fields once magnetic structures have expanded into the low-density, low-$\beta$ corona.  
This amounts to ignoring the data in the photosphere altogether in favor of a different, more idealized problem. 

The second approach involves generating initial thermodynamic states and flow fields by a dynamic relaxation process that holds the initial magnetic configuration fixed, 
and allows the model atmosphere to evolve to a dynamic state where non-magnetic forces are sufficient to prevent the Lorentz forces of the fixed magnetic field from disrupting structures over the time scale of observed photospheric evolution. 
Once this state is achieved, only then is it possible to apply PDFI electric fields in the photospheric layer and allow the system to evolve in a physical way.
This is a work in progress, and we will report on our results in a subsequent publication. 

\section{Visualizing Coronal Brightness with a Current-Based Spherical Emissivity Model}
\label{sec:emissivity}

In order to assess how well the data-driven simulations approximate the solar corona, comparisons to observational data, such as that from SDO AIA, must be made.  
For MHD simulations, thermodynamic variables can be convolved with the AIA filter response functions to provide a measure of coronal emission.  
However, for magneto-frictional simulations the only quantities available are the magnetic field and current density.  
One possibility in this case is to consider emission due to the dissipation of currents in the corona.
\citet{Cheung2012} found that the field line-averaged square of the current density served as an adequate high-temperature emission proxy for their Cartesian data-driven MF simulations, e.g. as illustrated in Figure~\ref{fig:j2AIA}.

To transition to spherical coordinates we developed the J2EMIS package,\footnote{\url{http://solartheory.ssl.berkeley.edu/cgi-bin/j2emis}} where we still use the average square of the current density as a proxy, but use a sampling methodology to calculate the emission, for both localized active region-scale and global simulations.
Results utilizing this sampling methodology are shown in the lower two panels of Figure~\ref{fig:j2ms}, where we have calculated the synthetic emission from the SMF magnetic field configuration of AR~11158 used to initialize the \radmhd simulation.
To facilitate integrating the emission, a Cartesian grid of the desired resolution is constructed to encompass the spherical data  and then rotated such that the x-axis is aligned with the line of sight of a chosen disk center.  
For each cell in this Cartesian grid a magnetic field line is traced from the center of the cell.
If the field line is closed (i.e., the field line intersects the model photosphere within some tolerance when traced in both directions) the square of the average current density over the length of the field line is determined and that value is saved as the emissivity of that cell.  
The integration of these emissivities is then calculated for the chosen line of sight and potentially the five other directions corresponding to the axes of the Cartesian grid. Calculation of the emissivity grid can be computationally expensive, depending on the desired resolution. A $600^3$ grid can use 3.5--7 GB of memory per process, and take 1--2 hours of wall-clock time on 96 processors.
We note that actual emission due to resistive heating depends on how much material is present to emit.  
Therefore, during the integration process we scale the emissivities with a radial profile based on the Baumbach-Allen density model \citep{Allen1947} to account for the density drop off with height in the corona.

The emission values produced with this methodology tend to have a limited dynamic range.  
Therefore using a simple log10-based luminance model to visualize the results ends up looking washed-out.  
More sophisticated high dynamic range luminance models must be used to see structure.  
We have had some success using two models. 
The first is the Schlick Uniform Rational Quantization method \citep{Schlick1995}.  
In this model a single parameter controls the non-linear brightness.  
In most cases this model is sufficient to bring out details in the emission structure.  
For cases where it is not, or where we want more control, we use the Reinhard/Devlin luminance model \citep{Reinhard2005}, 
which is a two-parameter model (for brightness and contrast) based upon photoreceptor physiology.  

\section{A Global MHD Model of the Corona and Heliosphere Driven by SMF Data}
\label{sec:global}

The global SMF model provides vector electric field values over the full Sun for an extended interval, but only out to a limited distance above the surface. Extending the model of evolving coronal and heliospheric conditions to greater heights, ultimately out to 1 AU, requires a time-dependent global MHD model like the one implemented for radial photospheric field measurements \citep{Hayashi2013}, but that can use the additional information computed from the global electric field provided by the CGEM global SMF model. We have adapted the existing global MHD model to use as input a time series of moderate-resolution global electric field data at 1.15~$R_\sun$, as described below. The model has been applied to data from an extended two-month global SMF simulation that includes the disk passage of AR~11158. The model can be applied to any extended interval for which the synoptic electric field is available, and it can be extended to heliocentric distances beyond 1 AU when necessary.

\begin{figure*}[thb]
\includegraphics[width=4.664in]{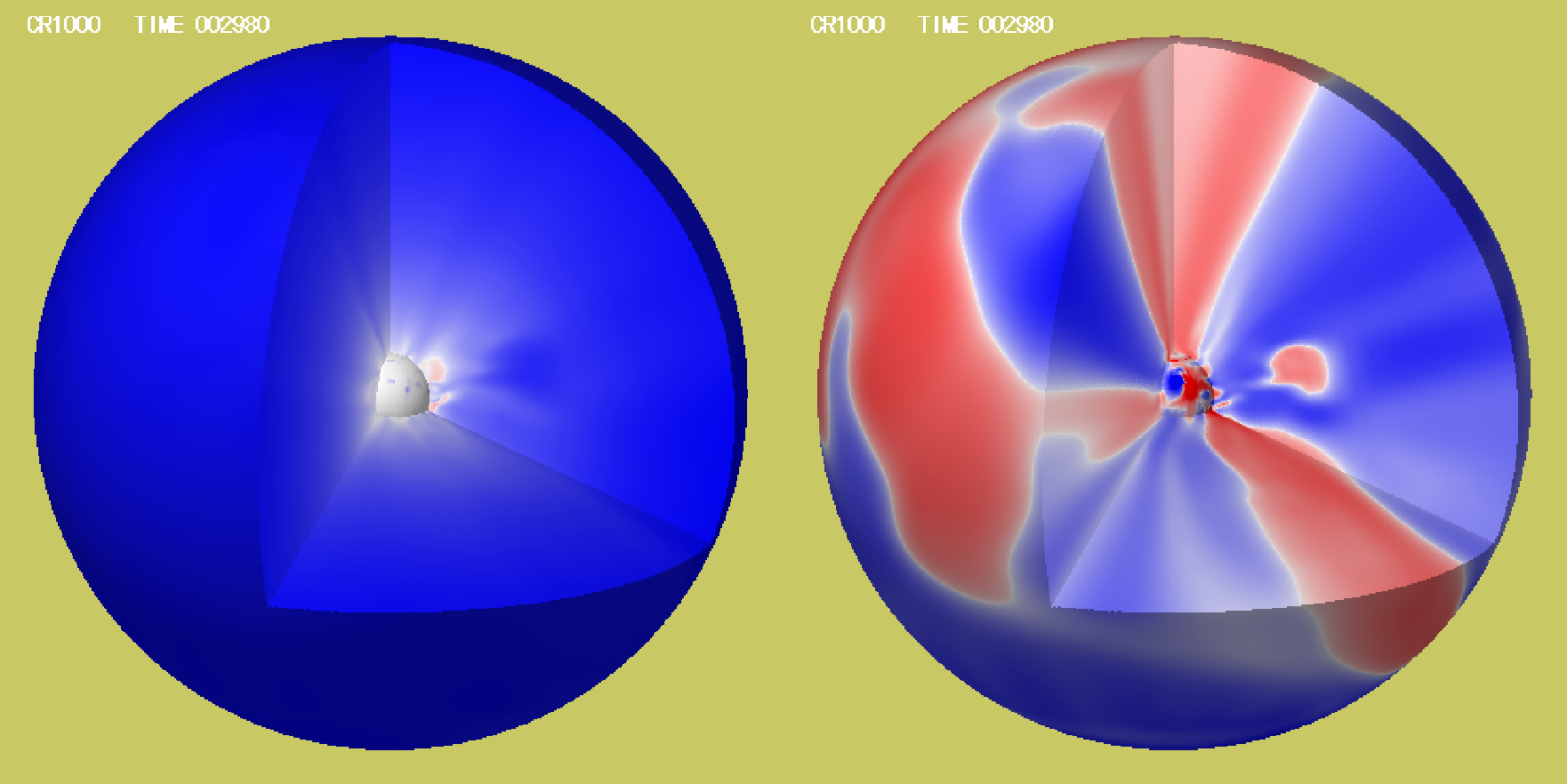} \includegraphics[width=2.336in]{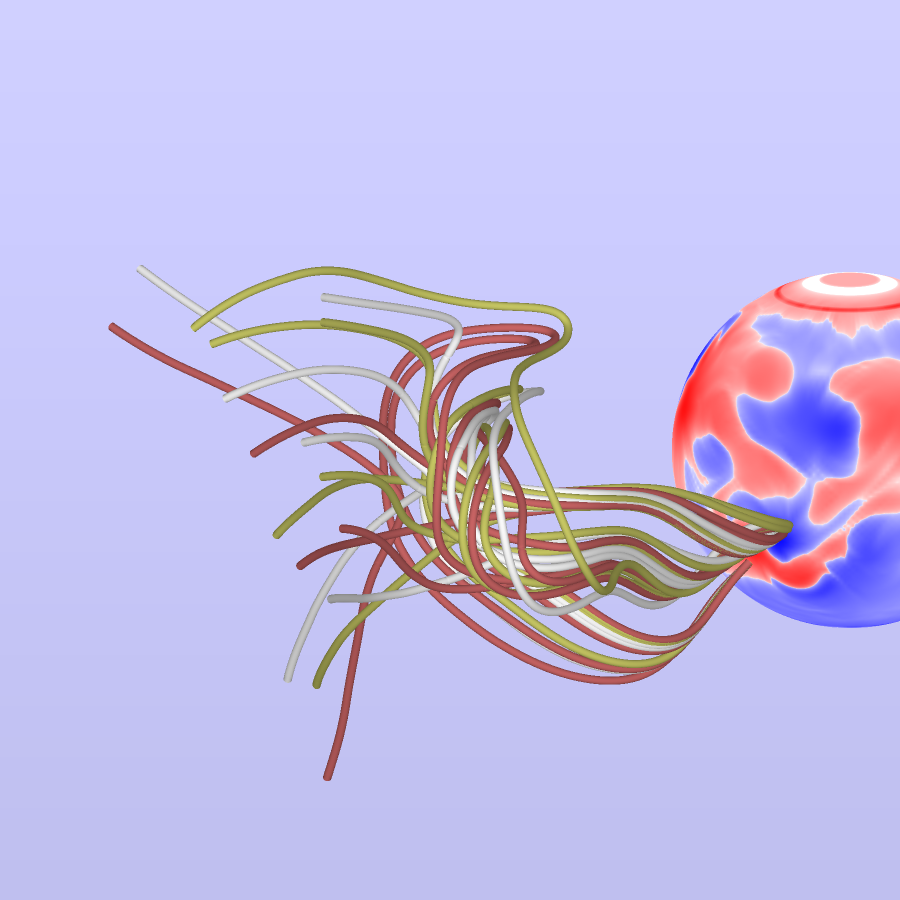}
\caption{ Left: Simulated plasma density in the Sun-to-Earth MHD model on selected cutout planes (two meridional planes and a plane at 20 degrees south), at $r \le 10\:R_\sun$.
Values are normalized by the density of the Parker solution; red, white, and blue show density ratios of 1.5, 1.2 and 1.0. 
Center: Radial magnetic field component $B_r$ normalized by the maximum absolute value at each radius. Blue (red) colors represent positive/outward (negative/inward) polarity.
Right: Selected field lines at $t \sim 2.5$\,d in the scale of SMF modeling, or $\sim 0.4$\,d after eruption of a twisted magnetic structure began.
The field lines are drawn for $1.15 \le r \le 6 \:R_\sun$.
Positive (negative) radial field $B_r$ on the bottom-boundary sphere at 1.15 $R_\sun$ is colored blue (red).
For visibility, the view point is 90 degrees West of the left and middle cross-section plots.
\label{fig:mhdglobal}}
\end{figure*}

We locate the interface boundary sphere at 1.15~$R_\sun$, and the magnetic field on the boundary sphere in the global Sun-to-Earth MHD model is directly driven using the SMF-derived electric field vector, as 
$c \vecE = -\partial_t\vecA$.

The interface boundary sphere in the CGEM framework is in the sub-Alfv\'{e}nic region; hence we need a proper numerical treatment to handle the incoming and outgoing wave modes.
We use the concept of projected normal characteristics \citep[e.g][]{Nakagawa1987} that offers a physics-based sub-Alfv\'{e}nic boundary treatment.
Details of the practical implementation of this method for global coronal modeling with time-varying boundary magnetic field are presented in \citet[][]{Hayashi2005,Hayashi2013} and \citet{Hayashi2018}.
In brief, the magnetic field vector is allowed to evolve arbitrarily, 
while the temporal variations of the other plasma variables (i.e., the density, temperature, and gas pressure, and the three components of plasma bulk flow) are determined using the normal projected characteristic method.
\citet{Hayashi2013} applied this method to a case where only the radial component $B_r$ and its evolution are specified. 

In the context of the CGEM framework, all three components of magnetic field and their temporal evolution are given by the lower-corona SMF model.
One of the principle challenges is setting up the simulation boundary treatment such that the three components of the simulated boundary $\vecB$ always match the three given components.
Another practical difficulty is that the SMF model does not provide information on plasma flow at the lower corona, which is required to complete the MHD equation system, or at least the induction equation.
The pseudo-plasma motion in the MF model is assumed to be parallel to the Lorentz force and hence perpendicular to the local magnetic field.
Therefore, we cannot use the global SMF plasma flows in a compressible MHD model, especially in coronal-hole or open-field regions in the global corona.

For the CGEM Sun-to-Earth MHD model, we choose to add the following steps to our earlier normal projected characteristic method to minimize the necessary development effort:
At each time step,
1) the boundary $\vecB$ is evolved tentatively in accordance with the electric field given from the global SMF model, while the temporal variations of the other variables (i.e., plasma density, temperature, and $\vecV$) are tentatively simulated in accordance with the normal projected characteristic method.
2) Adjustments to the simulated boundary variables are made depending on the value of the radial component of the plasma flow $v_r$, as follows:
(a) If the plasma is stagnant ($v_r=0$), all tentative simulated variables become final.
In practice, a looser criteria $|v_r|< 10\:m\, s^{-1}$ is used to determine whether a region is stagnant. 
(b) If the tentative $v_r$ is negative, $\vecV$ is forced to zero and the plasma density and temperature are adjusted according to the normal projected characteristic method.
(c) If $v_r$ is positive, all tentatively calculated temporal variations of the MHD variables are final.

In the first two cases, (a) and (b), upward- / outward-moving magnetic flux in the global SMF model enters into the domain of Sun-Earth global MHD simulation, without involving plasma motions; hence, the frozen-in condition is not preserved.
In the open-field coronal hole, with the last procedure (c), the plasma flows outward, parallel to the local boundary magnetic field.

The spherical grid of the global Sun-Earth MHD model is constructed to match the SMF model, and spans 128 by 256 points in the latitudinal and longitudinal directions, respectively.
In the current version, the heliocentric distance, $1.15 R_\sun \le r \le 1~AU$, is covered by 144 mesh elements with the size $\Delta r$ gradually varying.
OpenMP and MPI are implemented to achieve operational capability ---
using eight cores of a Xenon 3.7\,GHz CPU system, the Sun-Earth MHD model requires one day of wall-clock time to model coronal and interplanetary plasmas over one day of simulated evolution.

Figure \ref{fig:mhdglobal} provides a snapshot of an eruptive event obtained during an earlier global simulation of AR~11158.
The substantial amount of twisted magnetic flux present in the region makes it a good test bed for validating the modules coupling the global SMF model and Sun-to-Earth MHD model.
The left panel shows the excess coronal density and the center panel the radial direction of the magnetic field out to 10 $R_\sun$.
The panel on the right zooms in on the eruption of a twisted magnetic structure generated by the bottom-boundary electric field that began to emerge on the 1.15~$R_\sun$ sphere about 2.2\,d earlier.

The simulated eruption does not correspond to an actual solar event; hence, we do not claim accuracy for this simulation result.
Instead, we emphasize the promise of connecting the global SMF model, with its powerful capability for handling non-potential features in the lower corona, to the Sun-Earth MHD model, with its capability to numerically simulate twisting coronal magnetic features at various spatial scales and trace their coronal and interplanetary consequences. 
We expect such comprehensive modeling can be a foundation for further improvements and advances in operational space-weather modeling.

\section{Discussion and Conclusion}
\label{sec:conclusion}

We have presented an overview of the Coronal Global Evolutionary Model (CGEM) project to construct a framework for data-driven modeling to investigate the accumulation of magnetic energy that leads to eruptive events. 
The data driving is implemented through the PDFI electric field processing of the corrected HMI vector magnetograms, calibrated Doppler maps, and local correlation-tracking velocity fields. 
The time series of vector magnetograms and the resulting electric field maps make up the time-dependent boundary conditions that directly or indirectly drive the CGEM suite of numerical models:
the global surface flux transport (SFT) model, 
the spherical magnetofrictional (SMF) model, 
the active region-scale radiative magnetohydrodynamic (\radmhd) model, 
and the global corona--heliospheric MHD model. 
The CGEM MHD models were designed to start from either local or global SMF results and are driven by the same electric field formalism applied to the SMF simulation data. 
The PDFI electric fields have been made available to the community through the SDO Joint Science Operation Center (JSOC).

The development of the CGEM framework represents a significant advance for numerical modeling of the dynamic, time-dependent solar corona. 
The incorporation of temporal sequences of photospheric vector magnetic field observations into the boundary conditions of large-scale and global modeling represents the most direct data-driving approach to date. 
While there are still improvements to be made in the treatment of the atmospheric evolution and ensuring the self-consistency between the observed physical quantities and the time-evolution of the full MHD system, 
the CGEM deliverables now in place and available to the solar physics community enable systematic, quantitative investigation into the pre-eruption evolution and energization of active regions.  

Understanding the physics of and being able to reasonably estimate the storage and release of free magnetic energy in the solar corona is one of the main challenges for the prediction of \emph{where} and \emph{when} solar flares and coronal mass ejections occur -- which are some of the most important drivers of space weather for the Sun-to-Earth system.  

The observed photospheric evolution of ARs and among AR systems can vary tremendously. 
In general, photospheric signatures of magnetic flux emergence, magnetic flux cancellation, and both large-scale and localized shearing and twisting motions are observed in ARs that range in complexity from simple, isolated bipolar regions to complex delta-spot configurations. 
While there has been a concerted effort to identify common signatures, or even statistical trends, in the pre-flare/pre-eruption evolution of ARs, the fact is that often each of these signatures is present at some time, or even simultaneously, during the AR lifetime.    
The CGEM PDFI approach of converting HMI vector magnetogram sequences into time series of electric fields suitable for incorporation into various numerical models means that these flows and the resulting physical quantities, such as the fluxes of magnetic energy and relative magnetic helicity into the corona, are readily available for essentially every active region observed in the SDO era.
Additionally, utilizing the PDFI input in the data-driven SFT and SMF modeling yields critical information about the three-dimensional structure of these energized magnetic flux systems and the distribution of magnetic stress, electric currents, and other measures of non-potentiality for the source regions of flares and CMEs.  

The next major milestone in being able to predict --- and realistically model --- solar flares and CMEs is to understand their initiation process or processes. 
The rapid transition of sheared, twisted, and otherwise energized AR fields from a quasi-stable, quasi-equilibrium state to an unstable, run-away configuration that either drives or is driven by magnetic reconnection (or both), remains an important and active area of heliophysics research. 
The CGEM PDFI framework for direct data-driving of numerical models of the solar corona is a critical component for furthering the theoretical development and understanding of AR stability and solar flare\-/\-CME onset. 
The SMF approach can identify the regions of concentrated shear/\-twist/\-non-potentiality and quasi-unstable field regions in and around ARs. 
However, the diffusive ``frictional'' relaxation means the steep magnetic field gradients and strongly localized current densities required for the onset of fast magnetic reconnection are not resolved.
Thus a more complete physical model, i.e. the full MHD system, is necessary to resolve these structures and capture the impulsive nature of the onset and rapid reconfiguration of magnetic flux during eruption that converts the stored magnetic free energy into electromagnetic radiation, kinetic energy of the Alfv\'{e}nic reconnection jets and erupting material, particle acceleration, and bulk plasma heating. 

The PDFI electric field-driving of the full MHD system, which ensures the model's physical consistency, continues to require further development, but also represents a necessary and promising avenue of future research.
The 3D MHD evolution of the magnetic field state from the energized SMF configuration is one way to test the magnetofrictional instability thresholds for various ARs and determine just how important resolving the field/\-electric-current gradients are in determining the dynamics and evolution of the eruption onset and stable-to-unstable transitions. 
With the CGEM framework, we are now able to address a very interesting question:
Are the observed photospheric evolution (flows, emergence, cancellation) and the resulting estimate of energy accumulation \emph{sufficient} to account for the observed energy release of a flare or CME event? 
If so, we are well on our way to more realistic, physics-based modeling of the origin and evolution of energetic coronal transients. 
If not, there is something fundamental that our current observations and numerical models are missing. 
In either case, significant scientific progress can and will be made, our heliophysics modeling improved, and our space weather forecasting capabilities advanced.

\subsubsection*{Acknowledgements}
This work was supported by NASA and NSF through their funding of the CGEM project through NSF award AGS‐1321474 to UC Berkeley, NASA award 80NSSC18K0024 to Lockheed Martin, and NASA award NNX13AK39G to Stanford University.  
This work was also supported by NASA through the one-year extension to the CGEM project, ECGEM, through award 80NSSC19K0622 to UC Berkeley.  
Some data products created for CGEM are available at the Solar Dynamics Observatory Joint Science Operations Center (SDO JSOC) supported by NASA Contract NAS5-02139 (HMI) to Stanford University.
WPA, DJB, BJL, and XS were funded in part by NASA grant NNX17AI28G.
We wish to thank the US Taxpayers for their generous support for this project.

\bibliography{apj-jour,master}

\begin{thebibliography}{}
\expandafter\ifx\csname natexlab\endcsname\relax\def\natexlab#1{#1}\fi
\providecommand{\url}[1]{\href{#1}{#1}}
\providecommand{\dodoi}[1]{doi:~\href{http://doi.org/#1}{\nolinkurl{#1}}}
\providecommand{\doeprint}[1]{\href{http://ascl.net/#1}{\nolinkurl{http://ascl.net/#1}}}
\providecommand{\doarXiv}[1]{\href{https://arxiv.org/abs/#1}{\nolinkurl{https://arxiv.org/abs/#1}}}

\bibitem[{{Abbett}(2007)}]{Abbett2007}
{Abbett}, W.~P. 2007, \apj, 665, 1469, \dodoi{10.1086/519788}

\bibitem[{{Abbett} \& {Bercik}(2014)}]{Abbett2014}
{Abbett}, W.~P., \& {Bercik}, D.~J. 2014, in Meeting Abstracts, Vol. 224,
  American Astronomical Society, \#123.47 :
  \url{http://solarmuri.ssl.berkeley.edu/~abbett/public/Radmhd2S/}

\bibitem[{{Abbett} \& {Fisher}(2012)}]{Abbett2012}
{Abbett}, W.~P., \& {Fisher}, G.~H. 2012, \solphys, 277, 3,
  \dodoi{10.1007/s11207-011-9817-3}

\bibitem[{{Abbett} {et~al.}(2000){Abbett}, {Fisher}, \& {Fan}}]{Abbett2000}
{Abbett}, W.~P., {Fisher}, G.~H., \& {Fan}, Y. 2000, \apj, 540, 548,
  \dodoi{10.1086/309316}

\bibitem[{{Abbett} {et~al.}(2004){Abbett}, {Fisher}, {Fan}, \&
  {Bercik}}]{Abbett2004}
{Abbett}, W.~P., {Fisher}, G.~H., {Fan}, Y., \& {Bercik}, D.~J. 2004, \apj,
  612, 557, \dodoi{10.1086/422444}

\bibitem[{{Abbett} \& {Hawley}(1999)}]{Abbett1999}
{Abbett}, W.~P., \& {Hawley}, S.~L. 1999, \apj, 521, 906,
  \dodoi{10.1086/307576}

\bibitem[{{Allen}(1947)}]{Allen1947}
{Allen}, C.~W. 1947, \mnras, 107, 426, \dodoi{10.1093/mnras/107.5-6.426}

\bibitem[{{Cheung} \& {DeRosa}(2012)}]{Cheung2012}
{Cheung}, M.~C.~M., \& {DeRosa}, M.~L. 2012, \apj, 757, 147,
  \dodoi{10.1088/0004-637X/757/2/147}

\bibitem[{{Cheung} {et~al.}(2015){Cheung}, {De Pontieu}, {Tarbell}, {Fu},
  {Tian}, {Testa}, {Reeves}, {Mart{\'\i}nez-Sykora}, {Boerner}, {W{\"u}lser},
  {Lemen}, {Title}, {Hurlburt}, {Kleint}, {Kankelborg}, {Jaeggli}, {Golub},
  {McKillop}, {Saar}, {Carlsson}, \& {Hansteen}}]{Cheung2015b}
{Cheung}, M. C.~M., {De Pontieu}, B., {Tarbell}, T.~D., {et~al.} 2015, \apj,
  801, 83, \dodoi{10.1088/0004-637X/801/2/83}

\bibitem[{{Chintzoglou} {et~al.}(2019){Chintzoglou}, {Zhang}, {Cheung}, \&
  {Kazachenko}}]{Chintzoglou2019}
{Chintzoglou}, G., {Zhang}, J., {Cheung}, M. C.~M., \& {Kazachenko}, M. 2019,
  \apj, 871, 67, \dodoi{10.3847/1538-4357/aaef30}

\bibitem[{{Evans} \& {Hawley}(1988)}]{evans1988}
{Evans}, C.~R., \& {Hawley}, J.~F. 1988, \apj, 332, 659, \dodoi{10.1086/166684}

\bibitem[{Fisher(2020)}]{Fisher2020}
Fisher, G.~H. 2020, {Input files for test programs using the PDFI\_SS software
  library}, 0,  Zenodo, \dodoi{10.5281/zenodo.3711096}

\bibitem[{{Fisher} {et~al.}(2012){Fisher}, {Bercik}, {Welsch}, \&
  {Hudson}}]{Fisher2012a}
{Fisher}, G.~H., {Bercik}, D.~J., {Welsch}, B.~T., \& {Hudson}, H.~S. 2012,
  \solphys, 277, 59, \dodoi{10.1007/s11207-011-9907-2}

\bibitem[{Fisher {et~al.}(2020)Fisher, Kazachenko, Welsch, \&
  Lumme}]{Fisher2020pdfi}
Fisher, G.~H., Kazachenko, M.~D., Welsch, B.~T., \& Lumme, E. 2020, The
  PDFI\_SS Electric Field Inversion Software, 2020-03-15-44f7dd47cd,  Zenodo,
  \dodoi{10.5281/zenodo.3711571}

\bibitem[{{Fisher} {et~al.}(2020){Fisher}, {Kazachenko}, {Welsch}, {Sun},
  {Lumme}, {Bercik}, {DeRosa}, \& {Cheung}}]{Fisher2019}
{Fisher}, G.~H., {Kazachenko}, M.~D., {Welsch}, B.~T., {et~al.} 2020, \apjs,
  248, 2, \dodoi{10.3847/1538-4365/ab8303}

\bibitem[{{Fisher} {et~al.}(2010){Fisher}, {Welsch}, {Abbett}, \&
  {Bercik}}]{Fisher2010}
{Fisher}, G.~H., {Welsch}, B.~T., {Abbett}, W.~P., \& {Bercik}, D.~J. 2010,
  \apj, 715, 242, \dodoi{10.1088/0004-637X/715/1/242}

\bibitem[{{Fisher} {et~al.}(2015){Fisher}, {Abbett}, {Bercik}, {Kazachenko},
  {Lynch}, {Welsch}, {Hoeksema}, {Hayashi}, {Liu}, {Norton}, {Dalda}, {Sun},
  {DeRosa}, \& {Cheung}}]{Fisher2015}
{Fisher}, G.~H., {Abbett}, W.~P., {Bercik}, D.~J., {et~al.} 2015, Space
  Weather, 13, 369, \dodoi{10.1002/2015SW001191}

\bibitem[{{Goodman}(2012)}]{Goodman2012}
{Goodman}, M.~L. 2012, \apj, 757, 188, \dodoi{10.1088/0004-637X/757/2/188}

\bibitem[{{Hayashi}(2005)}]{Hayashi2005}
{Hayashi}, K. 2005, \apjs, 161, 480, \dodoi{10.1086/491791}

\bibitem[{{Hayashi}(2013)}]{Hayashi2013}
---. 2013, Journal of Geophysical Research (Space Physics), 118, 6889,
  \dodoi{10.1002/2013JA018991}

\bibitem[{{Hayashi} {et~al.}(2018){Hayashi}, {Feng}, {Xiong}, \&
  {Jiang}}]{Hayashi2018}
{Hayashi}, K., {Feng}, X., {Xiong}, M., \& {Jiang}, C. 2018, \apj, 855, 11,
  \dodoi{10.3847/1538-4357/aaacd8}

\bibitem[{{Hoeksema} {et~al.}(2014){Hoeksema}, {Liu}, {Hayashi}, {Sun},
  {Schou}, {Couvidat}, {Norton}, {Bobra}, {Centeno}, {Leka}, {Barnes}, \&
  {Turmon}}]{Hoeksema2014}
{Hoeksema}, J.~T., {Liu}, Y., {Hayashi}, K., {et~al.} 2014, \solphys, 289,
  3483, \dodoi{10.1007/s11207-014-0516-8}

\bibitem[{{Illenseer} \& {Duschl}(2009)}]{Illenseer2009}
{Illenseer}, T.~F., \& {Duschl}, W.~J. 2009, Computer Physics Communications,
  180, 2283, \dodoi{http://dx.doi.org/10.1016/j.cpc.2009.07.016}

\bibitem[{{Jiang} {et~al.}(2014){Jiang}, {Hathaway}, {Cameron}, {Solanki},
  {Gizon}, \& {Upton}}]{JiangJ2014}
{Jiang}, J., {Hathaway}, D.~H., {Cameron}, R.~H., {et~al.} 2014, \ssr, 186,
  491, \dodoi{10.1007/s11214-014-0083-1}

\bibitem[{{Kazachenko} {et~al.}(2014){Kazachenko}, {Fisher}, \&
  {Welsch}}]{Kazachenko2014}
{Kazachenko}, M.~D., {Fisher}, G.~H., \& {Welsch}, B.~T. 2014, \apj, 795, 17,
  \dodoi{10.1088/0004-637X/795/1/17}

\bibitem[{{Kazachenko} {et~al.}(2015){Kazachenko}, {Fisher}, {Welsch}, {Liu},
  \& {Sun}}]{Kazachenko2015}
{Kazachenko}, M.~D., {Fisher}, G.~H., {Welsch}, B.~T., {Liu}, Y., \& {Sun}, X.
  2015, \apj, 811, 16, \dodoi{10.1088/0004-637X/811/1/16}

\bibitem[{{Kissmann} \& {Pomoell}(2012)}]{Kissmann2012}
{Kissmann}, R., \& {Pomoell}, J. 2012, SIAM Journal on Scientific Computing,
  34, A763, \dodoi{10.1137/110834329}

\bibitem[{{Knoll} \& {Keyes}(2004)}]{Knoll2004}
{Knoll}, D.~A., \& {Keyes}, D.~E. 2004, Journal of Computational Physics, 193,
  357, \dodoi{10.1016/j.jcp.2003.08.010}

\bibitem[{{Komm} {et~al.}(1993{\natexlab{a}}){Komm}, {Howard}, \&
  {Harvey}}]{Komm1993a}
{Komm}, R.~W., {Howard}, R.~F., \& {Harvey}, J.~W. 1993{\natexlab{a}},
  \solphys, 145, 1, \dodoi{10.1007/BF00627979}

\bibitem[{{Komm} {et~al.}(1993{\natexlab{b}}){Komm}, {Howard}, \&
  {Harvey}}]{Komm1993b}
---. 1993{\natexlab{b}}, \solphys, 147, 207, \dodoi{10.1007/BF00690713}

\bibitem[{{Kurganov} \& {Levy}(2000)}]{Kurganov2000}
{Kurganov}, A., \& {Levy}, D. 2000, SIAM Journal on Scientific Computing, 22,
  1461, \dodoi{10.1137/S1064827599360236}

\bibitem[{{Leake} {et~al.}(2012){Leake}, {Lukin}, {Linton}, \&
  {Meier}}]{Leake2012}
{Leake}, J.~E., {Lukin}, V.~S., {Linton}, M.~G., \& {Meier}, E.~T. 2012, \apj,
  760, 109, \dodoi{10.1088/0004-637X/760/2/109}

\bibitem[{{Linker} {et~al.}(2001){Linker}, {Lionello}, {Miki{\'c}}, \&
  {Amari}}]{Linker2001}
{Linker}, J.~A., {Lionello}, R., {Miki{\'c}}, Z., \& {Amari}, T. 2001, \jgr,
  106, 25165, \dodoi{10.1029/2000JA004020}

\bibitem[{{Lionello} {et~al.}(2001){Lionello}, {Linker}, \&
  {Miki{\'c}}}]{Lionello2001}
{Lionello}, R., {Linker}, J.~A., \& {Miki{\'c}}, Z. 2001, \apj, 546, 542,
  \dodoi{10.1086/318254}

\bibitem[{{Low}(2010)}]{Low2010}
{Low}, B.~C. 2010, \solphys, 266, 277, \dodoi{10.1007/s11207-010-9619-z}

\bibitem[{{Lumme} {et~al.}(2019){Lumme}, {Kazachenko}, {Fisher}, {Welsch},
  {Pomoell}, \& {Kilpua}}]{Lumme2019}
{Lumme}, E., {Kazachenko}, M.~D., {Fisher}, G.~H., {et~al.} 2019, \solphys,
  294, 84, \dodoi{10.1007/s11207-019-1475-x}

\bibitem[{{Mart{\'{\i}}nez-Sykora} {et~al.}(2013){Mart{\'{\i}}nez-Sykora}, {De
  Pontieu}, {Leenaarts}, {Pereira}, {Carlsson}, {Hansteen}, {Stern}, {Tian},
  {McIntosh}, \& {Rouppe van der Voort}}]{MartinezSykora2013}
{Mart{\'{\i}}nez-Sykora}, J., {De Pontieu}, B., {Leenaarts}, J., {et~al.} 2013,
  \apj, 771, 66, \dodoi{10.1088/0004-637X/771/1/66}

\bibitem[{{Mok} {et~al.}(2005){Mok}, {Miki{\'c}}, {Lionello}, \&
  {Linker}}]{Mok2005}
{Mok}, Y., {Miki{\'c}}, Z., {Lionello}, R., \& {Linker}, J.~A. 2005, \apj, 621,
  1098, \dodoi{10.1086/427739}

\bibitem[{{Nakagawa} {et~al.}(1987){Nakagawa}, {Hu}, \& {Wu}}]{Nakagawa1987}
{Nakagawa}, Y., {Hu}, Y.~Q., \& {Wu}, S.~T. 1987, \aap, 179, 354

\bibitem[{{Pesnell} {et~al.}(2012){Pesnell}, {Thompson}, \&
  {Chamberlin}}]{Pesnell2012}
{Pesnell}, W.~D., {Thompson}, B.~J., \& {Chamberlin}, P.~C. 2012, \solphys,
  275, 3, \dodoi{10.1007/s11207-011-9841-3}

\bibitem[{{Reinhard} \& {Devlin}(2005)}]{Reinhard2005}
{Reinhard}, E., \& {Devlin}, K. 2005, IEEE Transactions on Visualization and
  Computer Graphics, 11, 13, \dodoi{10.1109/TVCG.2005.9}

\bibitem[{{Rogers}(2000)}]{Rogers2000}
{Rogers}, F.~J. 2000, Physics of Plasmas, 7, 51, \dodoi{10.1063/1.873815}

\bibitem[{{Scherrer} {et~al.}(2012){Scherrer}, {Schou}, {Bush}, {Kosovichev},
  {Bogart}, {Hoeksema}, {Liu}, {Duvall}, {Zhao}, {Title}, {Schrijver},
  {Tarbell}, \& {Tomczyk}}]{Scherrer2012}
{Scherrer}, P.~H., {Schou}, J., {Bush}, R.~I., {et~al.} 2012, \solphys, 275,
  207, \dodoi{10.1007/s11207-011-9834-2}

\bibitem[{{Schlick}(1995)}]{Schlick1995}
{Schlick}, C. 1995, in Photorealistic Rendering Techniques, ed. G.~{Sakas},
  S.~{M{\"u}ller}, \& P.~{Shirley} (Berlin, Heidelberg: Springer Berlin
  Heidelberg), 7, \dodoi{10.1007/978-3-642-87825-1_2}

\bibitem[{{Schou} {et~al.}(2012){Schou}, {Scherrer}, {Bush}, {Wachter},
  {Couvidat}, {Rabello-Soares}, {Bogart}, {Hoeksema}, {Liu}, {Duvall}, {Akin},
  {Allard}, {Miles}, {Rairden}, {Shine}, {Tarbell}, {Title}, {Wolfson},
  {Elmore}, {Norton}, \& {Tomczyk}}]{Schou2012}
{Schou}, J., {Scherrer}, P.~H., {Bush}, R.~I., {et~al.} 2012, \solphys, 275,
  229, \dodoi{10.1007/s11207-011-9842-2}

\bibitem[{{Schuck}(2008)}]{Schuck2008}
{Schuck}, P.~W. 2008, \apj, 683, 1134, \dodoi{10.1086/589434}

\bibitem[{{Stone} \& {Norman}(1992)}]{Stone1992a}
{Stone}, J.~M., \& {Norman}, M.~L. 1992, \apjs, 80, 753, \dodoi{10.1086/191680}

\bibitem[{{Sun}(2014)}]{sun2014}
{Sun}, X. 2014, arXiv e-prints, arXiv:1405.7353.
\newblock \doarXiv{1405.7353}

\bibitem[{{Sun} {et~al.}(2017){Sun}, {Hoeksema}, {Liu}, {Kazachenko}, \&
  {Chen}}]{sun2017}
{Sun}, X., {Hoeksema}, J.~T., {Liu}, Y., {Kazachenko}, M., \& {Chen}, R. 2017,
  \apj, 839, 67, \dodoi{10.3847/1538-4357/aa69c1}

\bibitem[{{Toriumi} {et~al.}(2020){Toriumi}, {Takasao}, {Cheung}, {Jiang},
  {Guo}, {Hayashi}, \& {Inoue}}]{Toriumi2020}
{Toriumi}, S., {Takasao}, S., {Cheung}, M. C.~M., {et~al.} 2020, \apj, 890,
  103, \dodoi{10.3847/1538-4357/ab6b1f}

\bibitem[{{van Leer}(1977)}]{vanleer1977b}
{van Leer}, B. 1977, J.\ Comp.\ Phys., 23, 276,
  \dodoi{10.1016/0021-9991(77)90095-X}

\bibitem[{{Welsch} {et~al.}(2013){Welsch}, {Fisher}, \& {Sun}}]{Welsch2013}
{Welsch}, B.~T., {Fisher}, G.~H., \& {Sun}, X. 2013, \apj, 765, 98,
  \dodoi{10.1088/0004-637X/765/2/98}

\bibitem[{{Wiegelmann} \& {Sakurai}(2012)}]{wiegelmann2012}
{Wiegelmann}, T., \& {Sakurai}, T. 2012, Living Reviews in Solar Physics, 9, 5,
  \dodoi{10.12942/lrsp-2012-5}

\bibitem[{{Young} {et~al.}(2003){Young}, {Del Zanna}, {Landi}, {Dere}, {Mason},
  \& {Landini}}]{Young2003}
{Young}, P.~R., {Del Zanna}, G., {Landi}, E., {et~al.} 2003, \apjs, 144, 135,
  \dodoi{10.1086/344365}

\end{thebibliography}


\end{document}